\documentclass[fleqn,usenatbib]{mnras}

\usepackage{newtxtext,newtxmath}
\usepackage[T1]{fontenc}
\usepackage{orcidlink}

\DeclareRobustCommand{\VAN}[3]{#2}
\let\VANthebibliography\thebibliography
\def\thebibliography{\DeclareRobustCommand{\VAN}[3]{##3}\VANthebibliography}

\usepackage{graphicx}	
\usepackage{amsmath}	
\graphicspath{{./}{figures/}}
\usepackage{float}
\usepackage[utf8]{inputenc}
\usepackage{xcolor}
\usepackage{graphicx}
\usepackage{amsmath}
\usepackage{ dsfont }
\usepackage{derivative}
\usepackage{float}
\usepackage{comment}
\usepackage{cuted}
\usepackage[T1]{fontenc}

\usepackage{mathtools,cuted}
\numberwithin{equation}{section}

\newcommand{\secref}[1]{Sec.~\ref{#1}}
\newcommand{\figref}[1]{Fig.~\ref{#1}}
\newcommand{\equnref}[1]{Eq.~\eqref{#1}}

\title[Inferring the role of binary neutron stars in r-process nucleosynthesis]{Inferring the role of binary neutron star mergers in r-process nucleosynthesis with multi-messenger observations using Cosmic Explorer and Einstein Telescope}

\author[A. Agarwal et al.]{
Aman Agarwal,$^{1}$\thanks{E-mail: aman.agarwal@uni-greifswald.de}\orcidlink{0000-0002-8685-5477}
Suvodip Mukherjee,$^{2}$ \orcidlink{0000-0002-3373-5236}
Daniel M.~Siegel$^{1}$ \orcidlink{0000-0001-6374-6465}
\\
$^{1}$Institute of Physics, University of Greifswald, 17489 Greifswald, Germany\\
$^{2}$Department of Astronomy \& Astrophysics, Tata Institute of Fundamental Research, 1, Homi Bhabha Road, Colaba, Mumbai 400005, India\\
}

\date{Accepted XXX. Received YYY; in original form ZZZ}

\pubyear{2026}

\begin{document}
\label{firstpage}
\pagerange{\pageref{firstpage}--\pageref{lastpage}}
\maketitle

\begin{abstract}
Identifying the cosmic origin of rapid neutron-capture (r-process) elements remains an open problem. Binary neutron-star (BNS) mergers and rare classes of core-collapse supernovae (CCSNe) represent the main contenders as major r-process production sites. 
Although BNS mergers could exclusively account for r-process nucleosynthesis, results from chemical evolution studies taking into account their delays with respect to star formation, observed BNS rates by gravitational-wave (GW) detectors, as well as issues with retention in low-mass halos suggest otherwise. Here, we propose a method to measure the contribution of BNS mergers to cosmic r-process nucleosynthesis with the third-generation GW detectors Cosmic Explorer and Einstein Telescope. It exploits the redshift-dependent correlation between the total number of BNS GW events and the average r-process abundances at redshifts $z \lesssim 1$. We apply this correlation technique to mock GW and abundance data, accounting for expected observational uncertainties in two limiting scenarios: GW events with electromagnetic counterpart (multi-messenger `bright-sirens') and without (`dark-sirens'). Using Fisher forecasts, we demonstrate that the fractional cumulative contribution of BNS mergers to the total cosmic r-process $F_{\rm{BNS,z0}}$ can be estimated to the $\lesssim 5-6\%$ precision level for both scenarios at $1\sigma$ for fiducial astrophysical scenarios with $F_{\rm{BNS,z0}} \gtrsim 0.1-1$. Furthermore, the method also yields estimates of the BNS delay-time distribution parameters comparable to other approaches. Although cosmic r-process abundances may be reconstructed from local observations at low metallicity, this method also provides a science case to identify signatures of neutron-capture elements beyond the local Universe.

\end{abstract}

\begin{keywords}
 nuclear reactions, nucleosynthesis, abundances -- neutron star mergers -- gravitational waves -- supernovae: general -- galaxies:abundances
\end{keywords}



\section{Introduction} 
\label{sec:intro}

The cosmic origin of rapid neutron-capture (r-process; \citealt{burbidge_synthesis_1957,cameron_origin_1957}) elements---roughly half of the elements heavier than iron---remains a fundamental open question in astrophysics and cosmology \citep{cowan_origin_2021,siegel_r-process_2022}. Several lines of evidence, ranging from measurements of radioactive isotopes on the sea floor \citep{wallner_abundance_2015,hotokezaka_short-lived_2015} to observed abundances of metal-poor stars in the Milky Way halo \citep{macias_stringent_2018} and those formed in the smallest dwarf galaxies \citep{ji_r-process_2016,tsujimoto_enrichment_2017}, suggest that the dominant site(s) of the r-process are prolific events much rarer than ordinary core-collapse supernovae (CCSNe), both in the early history of our Galaxy and today. The candidate sites as major contributors most widely discussed are neutron-star mergers \citep{lattimer_black-hole-neutron-star_1974,symbalisty_neutron_1982,eichler_nucleosynthesis_1989,freiburghaus_r-process_1999} and rare channels of CCSNe, such as those giving birth to a rapidly spinning magnetar (``magnetorotational SNe''; \citealt{leblanc_numerical_1970,symbalisty_expanding_1985,cameron_nucleosynthesis_2003,2006_Nishimura_MHD_SNe,metzger_proto-neutron_2007,winteler_magnetorotationally_2012}) and their more energetic cousins, those leading to a hyper-accreting black hole (``collapsars''; \citealt{macfadyen_collapsars_1999,pruet_nucleosynthesis_2003,surman_nucleosynthesis_2006,2019Nature_Siegel,siegel_super-kilonovae_2022}).

Whereas the first observed binary neutron star (BNS) merger GW170817 \citep{GW170817_discovery_paper} provided ample evidence that such mergers can execute an r-process, observed via the associated quasi-thermal kilonova emission (see \citealt{metzger_kilonovae_2020,siegel_gw170817_2019,margutti_first_2021,pian_mergers_2021} for reviews on the interpretation of the event), a similar signature has not yet been detected from isolated broad-lined Type Ic SNe (thought to be generated by magnetorotational SNe) or those observed in coincidence with long gamma-ray bursts (GRBs; thought to originate in collapsars). First observational searches for such signatures based on photospheric emission are inconclusive \citep{anand_collapsars_2024,rastinejad_hubble_2024,blanchard_jwst_2024}. This fact is not necessarily constraining, since r-process material and its photospheric as well as spectroscopic signatures can easily be buried behind several solar masses of ordinary SN ejecta \citep{2019Nature_Siegel,barnes_hydrodynamic_2023}. 

Arguments based on the chemical evolution history of the early Milky Way galaxy have been made in favour of both rare SNe (magnetorotational SNe and collapsars, hereafter collectively labelled rCCSNe; \citealt{2015_Cescutti_NS_in_r_process_MW_halo,wehmeyer_galactic_2015,cote_neutron_2019,2019Nature_Siegel,vandevoort_neutron_2020,yamazaki_possibility_2022,brauer_collapsar_2021,chen_inference_2025,saleem_mergers_2025}) and mergers \citep{shen_history_2015-1,duggan_neutron_2018,macias_constraining_2019,bartos_early_2019,tarumi_evidence_2021} as early r-process sources. A critical difference between the two types of r-process sources is their relative delay with respect to star formation with which they enrich the interstellar medium (ISM) with heavy elements. Collapsars, supercollapsars as their more extreme versions \citep{siegel_super-kilonovae_2022,2026ApJ_Agarwal_Ignition_supercollapsars}, and magnetorotational SNe are promptly generated following the deaths of very massive stars on timescales of $\sim\!\text{Myr}$ and can thus naturally explain r-process enrichment at very small metallicities, as early as the first generation of stars. Empirically found to occur in small dwarf galaxies at low metallicity \citep{fruchter_long_2006-1}, collapsars provide a natural explanation for early r-process enrichment in ultra-faint dwarf galaxies, such as Reticulum II \citep{ji_r-process_2016}, and metal-poor stars in the Galactic halo \citep{brauer_collapsar_2021}. In contrast, due to their inspiral driven by gravitational-wave (GW) emission, neutron-star mergers follow a power-law delay-time distribution (DTD) of $t^{-b}$ relative to star formation, with its canonical steepness of $b\approx 1$ and the existence of a more rapidly merging subpopulation with $b\lesssim 2$ being a matter of debate \citep{2019_Beniamini_Piran_BNS_DTD,Zevin_dtd_sgrb_2022,2025_Maoz_Ehud_DTD_paper}. Such power-law DTDs are disfavoured by chemical evolution analyses of r-process abundances in the high-metallicity stars of the Galactic disk \citep{hotokezaka_neutron_2018,siegel_gw170817_2019,cote_neutron_2019,saleem_mergers_2025}. `Prompt' rCCSNe also avoid the (larger) systemic kicks of neutron-star binaries, which easily exceed the escape velocity of small halos, as well as the iron co-production of at least two supernovae per binary, which cast doubt on whether mergers are the dominant sources of r-process elements in low-metallicity environments, including dwarf galaxies and globular clusters \citep{bonetti_neutron_2019,zevin_can_2019,skuladottir_neutron-capture_2019,naidu_evidence_2022,simon_timing_2023,kirby_r-process_2023}. Moreover, leaving issues of delay times, systemic kicks etc. ~aside, the downward revision of neutron-star merger rates as inferred from the recently concluded fourth observation run of the LIGO-Virgo-KAGRA network of GW detectors \citep{2025_GWTC4_r_and_p} raises further uncertainty in the total contribution of BNS mergers to r-process nucleosynthesis in the Universe \citep{saleem_mergers_2025}.

In this paper, we propose a novel method to determine the contribution of BNS mergers to the total r-process production in the Universe. Since black-hole--neutron-star mergers are likely insignificant relative to BNS mergers in terms of r-process enrichment \citep{chen_relative_2021,saleem_mergers_2025}, here we solely focus on BNS systems. If BNS mergers contribute to the cosmic production of r-process elements in a major way, the redshift evolution of the number of BNS GW events and some observational tracer of r-process abundances are correlated. We explore these correlations and demonstrate how they can be measured from future data to provide a precise measurement of the fractional contribution of BNS mergers to the cosmic r-process. The method is inspired by the correlation technique of \cite{2021_Mukherjee_Azadeh} proposed to infer the delay-time parameters of binary black-hole mergers with respect to star formation from the correlation between electromagnetic observations of emission line galaxies and GW observations of binary black-hole mergers.

The proposed correlation method requires r-process abundance and BNS GW data out to redshifts $z\gtrsim0.1$, which are not available with current GW detectors and electromagnetic telescopes.
With third-generation (3G) detectors, such as Cosmic Explorer\footnote{https://cosmicexplorer.org/} (CE; \citealt{punturo_einstein_2010,2019_Cosmic_explorer_white_paper}) and the Einstein Telescope\footnote{https://einsteintelescope.eu/} (ET; \citealt{2020_Einstein_Telescope_white_paper}), 
the detection horizons for BNS mergers are expected to increase to $z \approx \mathcal{O}(1)$ \citep{2022_3G_detection_capabilities} and a typical source localisation within $10-100 \ \rm{deg}^2$ (\citealt{2018_BNS_detection_capability_CE_paper}) is expected with increased probability for multimessenger follow-up detections with electromagnetic telescopes \citep{2011GReGr_chassande_mottin_multimessenger_3g_Detectors,2022_multimessenger_3G_applications,2024arXiv_Hu_Pei-jin_prospects_multimessenger_3G_era}. Whereas tens to hundreds of electromagnetic counterparts to GW events are expected in the 3G era, the uncertainties related to the local merger rate of BNS, the neutron star equation of state and its mass distribution can vary these estimates by $\sim\!60\%$ \citep{2025A&A_Loffredo_3G_detector_multi_messenger_prsopects,2025A&A_Colombo_multi_messenger_observations_3G_era}. Furthermore, even though the kilonova detection horizon of future telescopes, such as the Nancy Grace Roman Space\footnote{https://science.nasa.gov/mission/roman-space-telescope/} \citep{2013_Spergel_wfirst,2019_akeson_wfirst}, will, in principle, be of the order of $z\sim\mathcal{O}(1)$ \citep{2022ApJChase_kilonova_detectability}, the fraction of detectable kilonovae depends on the telescope's field-of-view, cadence choices and wavelength sensitivity \citep{2022ApJChase_kilonova_detectability,2016ApJBartos_JWST,2019ApJ_Cowperthaite_LSST,2019PAS_ZTF_overview}. Another possibility for host-galaxy identification and redshift measurement is the follow-up of associated short GRBs in a fraction of all BNS mergers (those successfully generating a GRB jet that additionally points to us). Therefore, in the absence of guaranteed electromagnetic counterparts, we consider two limiting scenarios for GW detection with 3G detectors: GW events with and without detected electromagnetic counterpart (i.e.~a `bright' or `dark' siren case, respectively). The bright siren scenario leads to a more precise estimate of the redshift-dependent fractional contribution of BNS mergers to the total cosmic r-process, due to reduced errors in the inferred luminosity distance and source redshift for such multimessenger events.

Observing abundances of neutron-capture elements beyond the local Universe remains a challenge. High-resolution spectroscopy of individual nearby stars in the Milky Way, its halo, as well as in local dwarf galaxies and globular clusters has created large databases of neutron-capture abundances over a wide range of metallicity and environments \citep{2018_JINA_database,SAGA_database}. Based on a comparison with individual stellar spectra, integrated-light spectroscopy of unresolved stellar populations of globular clusters has demonstrated the feasibility of determining neutron-capture element abundances including the r-process element Eu for globular clusters around other galaxies, providing concrete results for globular clusters of the Large Magellanic Cloud \citep{2009ApJ_Colucci_M31_ILS,2012ApJ_Colucci_Gc_abundances,2013MNRAS_Sakari_ILS_spectra}. Significant efforts have been devoted to determining abundances of $\alpha$-elements up to iron---specifically the metallicity---of galaxies across a wide range of galaxy masses and redshifts, using spectroscopy of galaxy continuum emission (e.g. \citealt{1998ApJ_Heckman,2004ApJ_rix_spectral_modeling, 2006MNRAS_Crowther_on_the_reliability, 2012A&A_Sommariva_stellar_metallicity, 2019MNRAS_cullen_vandels}) and via the gas phase using optical emission lines from HII regions (e.g. oxygen line measurements in \citealt{2017Peimbert_nebular_spectroscopy, kewley_understanding_2019, 2019A&AR_Maiolino_de_re} and iron line measurements in \citealt{1999ApJ_izotov_heavy_element, 2006A&A_izotov_chemical_composition,2018MNRAS_izotov_J0811, 2020ApJ_Kojima_Extremely_Metal_poor}). Recent observations by James Webb Space Telescope \citep{2006SS_Gardner_JWST} have also increased the redshift horizon of such abundance measurements from emission lines out to $z \sim 10$ \citep{2023ApJ_Isobe_JWST_identification,2024A&A_Schaerer_N_emitter,2024MNRAS_JI_GA-NIFS,2024A&A_Marques-chaves_N_emitters,2025A&A_Curti_JADES_JWST_SFR}. Beyond `light' metals up to iron, not much is known regarding abundances of neutron-capture elements at cosmological distances. In the following, we assume the availability of an r-process abundance tracer of galaxies at redshifts $z\gtrsim 0.1$ by the 3G GW detector era in the mid-to-late 2030s, either directly through progress on continuum spectroscopy, gas-phase abundance inference, or, indirectly, by linking neutron-capture element abundances to other observables via galaxy evolution studies. Finally, as a last resort, our method can also be applied to r-process abundances extracted from observations of local stellar spectra across the observed metallicity range, under the assumption that the observed r-process abundances of old low-mass stars at low metallicities reflect the nucleosynthesis conditions at high redshift---a general premise of stellar archaeology. Nonetheless, our method provides motivation to identify signatures of neutron-capture elements in the electromagnetic emission of galaxies beyond the local Universe.

The paper is structured as follows. We introduce the basic method to estimate the fractional cumulative contribution $F_{\rm BNS}(z=0)$ of neutron-star mergers to the cosmic r-process as well as their delay-time behaviour relative to star formation from the correlations of their GW signals and r-process abundance data in \secref{sec:Framework}. In \secref{sec:GCE_modelling}, we model the dependence of these two observational quantities on $F_{\rm BNS}(z=0)$ and the parameters of the DTD of BNS mergers with respect to the star formation rate (SFR). We present a Fisher forecast of the precision with which $F_{\rm BNS}(z=0)$ and the DTD parameters can be extracted by creating a mock database of GW events expected in the 3G era and mock estimates of cosmological r-process abundances in \secref{sec:fisher Analysis}. Section \ref{sec:conclusion} provides further discussion and summarizes our conclusions.

\section{Correlation between Gravitational-Wave Sources and R-process Abundances}
\label{sec:Framework}
 
\subsection{Correlation Technique}
\label{sec:correlation_function_definition}

We assume that at least a fraction of the observed cosmic r-process material originates in BNS mergers, providing a certain degree of correlation between an observational tracer $\mathrm{Ab^{\rm El}}(z)$ of cosmic r-process abundances as a function of redshift and the redshift-dependent number $N_{\rm GW}(z)$ of observed GW events from BNS systems. We think of cosmic r-process abundances here as a yet to be defined average over mean r-process abundances of galaxies at a certain redshift (see, in particular, Appendix \ref{sec:GCE_evolution}). One may express the observed $N_{\rm GW}(z)$ and $\mathrm{Ab^{El}}(z)$ as functions $f$ and $g$ of the cosmic star formation history $\psi_{\rm SF}(z)$, respectively,
\begin{equation}
    N_{\rm GW} (z) = f[\psi_{\mathrm{SF}}(z)],\mskip20mu {\rm Ab}^{\rm El}(z) = g[\psi_{\mathrm{SF}}(z)]. \label{eq:correlation_technique_foundation1}
\end{equation}
For concreteness, here we take a tracer for europium ($\rm{El}=\rm{Eu}$), which is almost exclusively synthesized by the r-process in the solar system. However, the method can use other tracer elements $\rm{El}$ and can be extended to simultaneously monitor multiple tracer elements or can use an aggregate proxy of the r-process.

Provided that a unique relation between star-formation history and an r-process abundance proxy exists, the relations \eqref{eq:correlation_technique_foundation1} may be written as
\begin{equation}
    N_{\rm GW} (z) = f\circ g^{-1}[{\rm Ab}^{\rm Eu}(z)].
    \label{eq:abstract_corr}
\end{equation}
In this case, the relation between $N_{\rm GW}(z)$ and $\mathrm{Ab^{Eu}}(z)$ becomes independent of the star formation history. It is then most sensitive to the relative behaviour of how both observables follow the star formation history. 

We investigate this relationship by introducing the correlation function $K(\theta)$, defined via
\begin{equation}
    N_{\rm GW} (z) = K(\theta){\rm Ab}^{\rm Eu}(z),
    \label{eq:correlation_general_form}
\end{equation}
where the set of parameters $\theta$ capture, in particular, the behaviour of the observables relative to star formation, including the fraction $F_{\rm{BNS,z0}}$ of the BNS mergers' contribution to the total cosmic r-process up to $z=0$. Future observations to measure the redshift-dependent ratio $K (\theta) = N_{\rm GW}(z)/\mathrm{Ab^{Eu}}(z)$ can be used to estimate the parameters $\theta$ (including $F_{\mathrm{BNS,z0}}$), e.g. using Bayesian statistics.

\subsection{ Parameter Estimation} 
\label{subsec:data_covariance}

Future observational data $\hat{N}_{\mathrm{GW}}$ and $\hat{\rm{Ab}}^{\rm Eu}$ (henceforth, hats denote observational data) in a given redshift range $\Delta z$ around $z$ will be estimated  from independent and identically distributed (`iid', that is, derived from the same probability distribution in the particular redshift bin) observations of GWs from BNS and r-process abundances of galaxies, respectively. Correlation data $\hat{K}(z)$ will be obtained as the ratio of these two quantities. Given astrophysical models $K_{\mathrm{model}}(\theta,z)$, $\mathrm{Ab^{Eu}}_{\mathrm{model}}(\theta,z)$, and $N_{\mathrm{GW,model}}(\theta,z)$, which provide expected values of observed quantities as a function of the parameters $\theta$, we can define a likelihood function of observing a certain data value $\hat{K}(z)$ given parameter values $\theta$. Using this likelihood function, we can estimate the parameter values $\theta$ that best fit our observed correlations $\hat{K}(z)$.

From the central limit theorem, the `iid' quantities $N_{\mathrm{GW}}$ and $\mathrm{Ab^{Eu}}$, and their combination $K$, are expected to follow a multivariate normal distribution.  
Given observed values $\hat{N}_{\mathrm{GW}}(z)$ and $\hat{\mathrm{Ab}^{\rm Eu}}(z)$ as vectors of size $N$, with $N$ being the number of redshift bins $z_i$, $i=0,\ldots,N$, the log-likelihood of the parameters $\theta$ can be defined using normal distributions as
\begin{eqnarray}
    \log L  \propto &-& L_1 -L_2 \nonumber\\
     \equiv &-&\frac{1}{2} \left[ \hat K(z) - K_{\mathrm{model}}(\theta,z)\right]^{T}\mathds{C}^{-1}_1(z)\nonumber\\
    &&\times\left[ \hat K(z) - K_{\mathrm{model}}(\theta,z)\right]\nonumber\\
    &-&\frac{1}{2} \left[ \hat{\mathrm{Ab^{Eu}}}(z) - \mathrm{Ab^{Eu}}_{\mathrm{model}}(\theta,z)\right]^T\mathds{C}^{-1}_2(z)\nonumber\\
    &&\times\left[ \hat{\mathrm{Ab^{Eu}}}(z) - \mathrm{Ab^{Eu}}_{\mathrm{model}}(\theta,z)\right],
\label{eq:general_likelihood_gaussian_proposed}
\end{eqnarray}
where $\mathds{C}_1(z)$ and $\mathds{C}_2(z)$ are covariance matrices of dimensions $N\times N$, which take observational errors into account and are described in detail below. 

The first term $L_1$ of the likelihood $L$ helps estimate the parameters $\theta$ that best fit the observed correlation data $\hat{K}$, including parameters of the DTD of BNS mergers relative to star formation as well as the total contribution of BNS mergers to the cosmic r-process. However, the correlation is scale-invariant in the sense that multiplying both $N_{\mathrm{GW}}$ and $\mathrm{Ab^{Eu}}$ by a constant factor leaves the correlation function in Eq.~\eqref{eq:correlation_general_form} unchanged. By individually fitting for one of the two datasets in the second term $L_2$ of the likelihood $L$, we fix the absolute scale of the two datasets and exclude this degeneracy. The term $L_1$ containing the correlation data can constrain $F_{\rm{BNS,z0}}$ largely independently of assumptions related to the star formation history and other common astrophysical or cosmological uncertainties underlying both BNS merger rates and r-process production via BNS mergers (as motivated in Secs.~\ref{sec:correlation_function_definition} and \ref{sec:GCE_modelling}). The term $L_2$ further contributes to the precision in estimating $F_{\rm{BNS,z0}}$, as the abundance terms contained in $L_2$ are directly sensitive to changes in $F_{\rm BNS,z0}$. The above likelihood $L$ is constructed to only measure the correlation of cosmic r-process abundances with BNS mergers and the total fractional contribution to the latter caused by BNS. It is not designed to measure the local merger rate of BNS. Such a measurement would require an additional term in the likelihood involving $\hat{N}_{\rm GW}(z)$ -- a straightforward, possible extension of the current method.

The covariance matrices in  
Eq.~\eqref{eq:general_likelihood_gaussian_proposed} (i.e., for the given normal distributions) are defined as follows:
\begin{eqnarray}
    \mathds{C}_1 &=& \left\langle \left[\hat K(z) - K_{\mathrm{model}}(\theta,z)\right]\left[\hat K(z) - K_{\mathrm{model}}(\theta,z)\right]^T\right\rangle, \nonumber\\
    \label{eq:C1_general}\\
    \mathds{C}_2 &=&  \Big\langle\left[\hat{\mathrm{Ab}}^{\rm Eu}(z) - \mathrm{Ab^{Eu}}_{\mathrm{model}}(\theta,z)\right]\nonumber\\
    &&\times
    \left[ \hat{\mathrm{Ab}}^{\rm Eu}(z) - \mathrm{Ab^{Eu}}_{\mathrm{model}}(\theta,z)\right]^{T}\Big\rangle. \label{eq:C3_general}
\end{eqnarray}
The angle brackets denote the mean over an ensemble of data with different noise realizations around the expected value. Since multiple realizations of the total number of BNS GW events $\hat{N}_{\rm GW}(z)$ or the mean Eu abundances $\hat{\mathrm{Ab}}^{\rm Eu}(z)$ in a certain redshift bin are not possible, the covariance matrix must be approximated according to theoretical expectations. We expect individual GW and abundance observations (and their corresponding observational uncertainties) to be independently obtained, and hence, we do not expect a statistical correlation between observations in different redshift bins.  
Therefore, the above covariance matrices are of diagonal form, $\left(\mathds{C}_k\right)_{ij} = \delta_{ij}C_k(z_i)$ for $k=1,2$, and the likelihood function \eqref{eq:general_likelihood_gaussian_proposed} becomes
\begin{eqnarray}
    \log L  \propto  \sum_{z_i} &-& \frac{\left[ \hat K(z_i) - K_{\mathrm{model}}(\theta,z_i)\right]^2}{2C_1(z_i)}\nonumber\\
    &-& \frac{\left[ \hat{\mathrm{Ab^{Eu}}}(z_i) - \mathrm{Ab^{Eu}}_{\mathrm{model}}(\theta,z_i)\right]^2}{2C_2(z_i)}.
\label{eq:univ_likelihood_gaussian_proposed}
\end{eqnarray}

We model the diagonal elements of the covariance matrix using expected observational and instrumental uncertainties in each redshift bin as
\begin{eqnarray}
    C_1 (z_i) &=& \sigma^2_{\hat{K}-\hat{N}_{\mathrm{GW}}}(z_i)+ \sigma^2_{\hat{K}-z}(z_i) + \sigma^2_{\hat{K}-\mathrm{\hat{Ab}^{Eu}}}(z_i), \nonumber\\
    \label{eq:covariance_matrix}\\
    C_2(z_i) &=& \sigma^2_{\mathrm{\hat{Ab}^{Eu}}}(z_i)
    \label{eq:covariance_dprime_matrix}.
\end{eqnarray}
The covariance function $C_1$ comprises the uncertainties in measuring the number of GW sources ($\sigma^2_{\hat{K}-\hat{N}_{\mathrm{GW}}}$), the redshift ($\sigma^2_{\hat{K}-z}$), and the abundance tracer of the cosmic r-process ($\sigma^2_{\hat{K}-\mathrm{\hat{Ab}^{Eu}}}$), propagated to the uncertainty of $\hat{K}$ via the chain rule. The covariance function $C_2$ consists solely of the measurement uncertainties in the abundance tracer of the r-process ($\sigma^2_{\mathrm{\hat{Ab}^{Eu}}}$).

The total number of GW events in a redshift bin $\hat{N}_{\rm GW}$ has two sources of uncertainty: counting statistic errors and redshift uncertainty associated with each event. For the counting uncertainty, we approximate GW sources in a particular redshift range to occur with a fixed expected rate, i.e., the BNS rate $R_{\rm BNS,ML}(z)$ for which the likelihood \eqref{eq:general_likelihood_gaussian_proposed} maximizes. The approximation is justified as deviations from this expected rate are quadratically suppressed in the likelihood function by definition. Since the GW observations are independent, the counting statistic $\hat{N}_{\mathrm{GW}}$ with a mean rate of occurrence $R_{\rm BNS,ML}(z)$ in a particular redshift bin should follow a Poisson distribution with standard deviation $\sigma_{\hat{N}_{\mathrm{GW}}} = \sqrt{\hat{N}_{\mathrm{GW}}}$. Thus, the error in measuring the number of GW sources is propagated to the correlation data $\hat{K}$ as
\begin{equation}
    \sigma^2_{\hat{K}-\hat{N}_{\mathrm{GW}}}(z_i) = \left(\frac{\partial \hat{K}}{\partial z}\frac{\partial z}{\partial \hat{N}_{\mathrm{GW}}}\bigg|_{z_i}\right)^2\sigma^2_{\hat{N}_{\mathrm{GW}}}(z_i). \label{eq:grav_sources_errors}
\end{equation}

The redshift uncertainty associated with each GW event has to be modelled differently for BNS events with and without electromagnetic (EM) counterparts. Without an EM counterpart, GW observations alone can only provide the luminosity distance to the GW source, which can be converted to a redshift assuming a fixed cosmology. The observational error in luminosity distance for BNS
assuming a cosmological model with parameters denoted by $\alpha_c$, is translated to data as  
\begin{equation}
    \sigma^2_{\hat{K}-z}(z_i) = \left(\frac{\partial \hat{K}}{\partial z}\right)^2\left(\frac{\partial z}{\partial d_{\rm L}}\bigg|_{\alpha_c}\right)^2\sigma^2_{d_{\rm L}}(z_i),\\
    \label{eq:redshift_error}
\end{equation}
where the derivative $\frac{\partial z}{\partial d_{\rm L}}\big|_{\alpha_c}$ is the Jacobian of the map between luminosity distance and redshift, and $\sigma_{d_{\rm L}}$ is the luminosity distance measurement uncertainty. 
For BNS events with EM counterparts that lead to the identification of the host galaxy and thus of a redshift measurement, the uncertainty in determining $z$ is expected to shrink significantly\footnote{This method requires galaxy surveys up to redshifts of $z\approx1$.}. 
We assume it to be
\begin{equation}
    \sigma^2_{\hat{K} - z}(z_i) = \left(\frac{\partial \hat{K}}{\partial z}\right)^2 \sigma^2_{\rm spec}(z_i),\label{eq:redshift_error2}
\end{equation}
where $\sigma_{\rm spec}(z)\approx \sigma_0(1+z)$ is the spectroscopic redshift uncertainty, with $\sigma_0 = 10^{-4}$ based on spectroscopic redshift measurement errors (e.g. \citealt{2017_GW170817_spectro_survey,GW170817_optical_paper}).

The error in measuring $\mathrm{\hat{Ab}^{Eu}}$ is translated to $\hat{K}$ by 
\begin{equation}
     \sigma^2_{\hat{K}-\mathrm{\hat{Ab}^{Eu}}} =  \left(\frac{\partial \hat{K}(\theta)}{\partial z}\frac{\partial z}{\partial \mathrm{\hat{Ab}^{Eu}}}\right)^2 \sigma^2_{\mathrm{\hat{Ab}^{Eu}}} (z),\label{eq:abundance_error}
\end{equation}
where $\sigma_{\mathrm{\hat{Ab}^{Eu}}}$ is the uncertainty in the r-process abundance proxy measurement.

 \section{Illustration of the Correlation Technique}
 \label{sec:GCE_modelling}
 
In this section, we investigate expected correlations concretely by modelling $N_{\mathrm{GW}}$ and $\mathrm{Ab^{Eu}}$ and examining their correlation function and its sensitivity to the parameters $\theta$, including $F_{\rm{BNS,z0}}$. 
 
For an estimate of $N_{\mathrm{GW}}(z)$, we integrate the modelled BNS merger rate $R_{\mathrm{BNS}}(t;t_{\mathrm{min}},b)$ over comoving cosmic volume and the observation time. The rate $R_{\mathrm{BNS}}(t,t_{\mathrm{min}},b)$ as a function of cosmic time $t$ is assumed to follow the SFR with some delay due to the GW-driven binary inspiral time, 
\begin{eqnarray}
R_{\mathrm{BNS}}(t;t_{\mathrm{min}},b)=
c_{\mathrm{BNS}}\int_{t(z=10)}^{t} \mskip-40mu D_{\rm BNS}(t-t';t_{\mathrm{min}},b)  \nonumber\psi_{\mathrm{SF}}(t') \mathrm{d} t' ,\label{eq:rate_BNS} \\
\end{eqnarray}
where the constant $c_{\mathrm{BNS}}$ normalizes the expression (and thus $ N_{\rm GW,model}$) to the currently observed rate at redshift $z=0$ of $R_{\mathrm{BNS,z0}}=100\, \mathrm{Gpc}^{-3}\mathrm{yr}^{-1}$ \citep{2025_GWTC4_r_and_p}. We employ the cosmic SFR $\psi_{\mathrm{SF}}(t)$ of \citet{Madau_Fargos_2014} 
and take $z=10$ as the highest redshift of our model. 
The DTD $D(t;t_{\mathrm{min}},b)$ (see, e.g., \cite{2013A&A_hao_delay_time_sgrb_progenitor,2015_Wanderman_Piran_sGRB_pop_DTD,2018MNRAS_Anand_DTD_from_SGRB,2020_Adhikari_dtd_papers,Zevin_dtd_sgrb_2022,2022ApJ_Nugent_SgRB_host_galaxy_dtd} for an estimation of the DTD from existing observations) is given by a power law\footnote{As a fiducial value, one expects $b\simeq-1$, since the orbital separation $a$ of BNS progenitor systems is thought to be nearly constant in log space, $\mathrm{d}N/\mathrm{d}\log a \simeq \text{const.}$ \citep{2012_Sana_orbital_periods}, and the inspiral time scales as $t_{\mathrm{insp}}\propto a^4$ for quasi-circular binaries in linearised general relativity \citep{1964_Peters_t_propto_a_power_4_derivation}, which leads to $\mathrm{d}N/\mathrm{d}t_{\mathrm{insp}} \propto t_{\mathrm{insp}}^{-1}$. Due to the strong dependence $t_{\mathrm{insp}}\propto a^4$, the initial orbital separations must deviate strongly from the above mentioned observations, $\mathrm{d}N/\mathrm{d} a\propto a^{-(1+\gamma)}$ with $|\gamma|\gtrsim 2-4$ instead of $\gamma=0$ for the resulting $\mathrm{d}N/\mathrm{d}t_{\mathrm{insp}} \propto t_{\mathrm{insp}}^{-(1+\gamma/4)}$ to deviate significantly from $\propto t_{\mathrm{insp}}^{-1}$.} 
as,
\begin{equation}
    D(t) = \Theta(t-t_{\mathrm{min}})t^{b},
\label{eq:delay_time_definition}
\end{equation}
where $\Theta$ is the Heaviside function and $t_{\mathrm{min}}$ and $b<0$ are free parameters. Considering the lifetime of BNS progenitor stars, the minimum delay $t_{\rm min}$ relative to star formation with which BNS merge has a theoretical lower limit of at least $\approx\! 20$--$40$\,Myr. 

The event number $\mathrm{d} N_{\rm GW, model}$ of BNS in a differential comoving volume $\mathrm{d}V$ and for an observation time $t_{\rm obs}$ is calculated as 
 \begin{equation}
     \mathrm{d} N_{\rm GW, model} = \frac{R_{\rm BNS}}{1+z}  \mathrm{d}V \times t_{\mathrm{obs}}.
 \end{equation}
 Here, the factor $(1+z)^{-1}$ accounts for cosmological time dilation from source time to observer time. The differential comoving volume is translated into a differential redshift interval using $\mathrm{d}V / \mathrm{d}z =  c (1 + z)^2 D^2_A H(z)^{-1} \mathrm{d} \Omega$, with $\Omega$ being the solid angle, $ D_A$ the angular distance, $H$ the Hubble parameter, and $c$ the speed of light. In a given redshift interval $\mathrm{d}z$, one therefore has \begin{equation}
     \frac{\mathrm{d} N_{\rm GW, model}}{\mathrm{d}z} = \int \mathrm{d}\Omega \frac{R_{\rm BNS}}{1+z}  \frac{\mathrm{d}^2V}{\mathrm{d}z\mathrm{d}\Omega} \times t_{\mathrm{obs}}.
 \end{equation}
 Assuming isotropy, the integral $\int d\Omega$ results in a factor of $4\pi$. The observable BNS source population number $N_{\rm GW,model}$ as a function of redshift is thus calculated by integrating the event number $\mathrm{d} N_{\rm GW, model}$ over a certain redshift interval $[z- \Delta z/2, z+ \Delta z/2]$ around redshift $z$:
 \begin{eqnarray}
    N_{\rm GW,model}(z) = & \phi(z)
    \times 
    \int_{t(z-\Delta z/2)}^{t(z+ \Delta z/2)} 
     4\pi \frac{R_{\rm BNS}(t',t_{\mathrm{min}},b)}{1+z(t')} \nonumber \\
    & \frac{\mathrm{d}^2V}{\mathrm{d}z\mathrm{d}\Omega} \frac{\mathrm{d}z}{\mathrm{d}t'} \mathrm{d}t' \times t_{\mathrm{obs}},
\label{eq:N_GW}
\end{eqnarray}
where $\phi(z)$ is the GW event selection function, which is detector-specific (see Appendix~\ref{sec:Mock_GW_catalogue}) and $\mathrm{d}z/\mathrm{d}t =  H(z) (1+z)$. 
The factors $\frac{\mathrm{d}^2V}{\mathrm{d}z\mathrm{d}\Omega}$ and $\mathrm{d}z/\mathrm{d}t$ are evaluated as functions of redshift for the cosmological parameters obtained by the \citet{2018_Planck_cosmological_params}. The observation time is fixed to a fiducial value of $t_{\mathrm{obs}}=1$\,yr for the rest of the analysis.\footnote{The estimated uncertainties of the parameters are only expected to decrease as $t_{\mathrm{obs}}$ increases due to reduced combined errors of a large number of observations.}

To obtain an estimate for $\mathrm{Ab}^{\mathrm{Eu}}$ as a function of redshift, we model cosmic r-process enrichment with the cosmic chemical evolution (CCE) model described in Appendix~\ref{sec:GCE_evolution}. As the proxy quantity $\mathrm{Ab}^{\mathrm{Eu}}$, we take the usual europium abundance $\mathrm{[Eu/H]}(z)=\log[N_{\rm Eu}(z)/N_{\rm H}(z)]-\log(N_{\rm Eu}/N_{\rm H})_\odot$ (where $N$ refers to the number of atoms), offset by $+5$ to avoid mathematical pathologies when computing correlations with GW events. The cosmic Eu abundance $\mathrm{[Eu/H]}(z)$ in a certain redshift bin around $z$ is obtained by averaging the Eu content of individual galaxies across a range of masses with their specific star formation histories. Chemical evolution in each galaxy is modelled by a simple one-zone approach, in which both BNS and other astrophysical sources without delay relative to star formation (subsumed as rCCSNe here) contribute to r-process nucleosynthesis,
\begin{equation}
    {\mathrm{Ab}^{\mathrm{Eu}}}_{\mathrm{model}}(z(t),\theta) = \mathrm{Ab^{Eu}}_{\mathrm{BNS}}(t,\theta) + \mathrm{Ab^{Eu}}_{\mathrm{rCCSNe}}(t,\theta), 
    \label{eq:ab_final_model}
\end{equation}
where $\theta = (t_{\mathrm{min}},b,F_{\rm{BNS,z0}})$. The CCE model has several more parameters as defined in Appendix~\ref{sec:GCE_evolution}, and the set $\theta$ only refers to the parameters that we are interested in estimating. 

The fractional cumulative contribution $F_{\rm BNS}(z,\theta)$ of BNS to Eu production at different redshifts is given by
\begin{equation}
    F_{\rm BNS}(z(t),\theta) = \frac{\mathrm{Ab^{Eu}}_{\mathrm{BNS}}(t,\theta)}{\mathrm{Ab^{Eu}}_{\mathrm{BNS}}(t,\theta) + \mathrm{Ab^{Eu}}_{\mathrm{rCCSNe}}(t,\theta)}, \label{eq:F_BNS_reln}
\end{equation}
with $F_{\rm BNS}(z=0,\theta) = F_{\rm{BNS,z0}}$. For purposes of illustration in this section, we fix certain target values of $F_{\rm{BNS,z0}}$ to represent different possible astrophysical scenarios and find the corresponding products of volumetric rate and Eu-yield $M_{\mathrm{s}}$ (Eq.~\eqref{eq:M_s}) for $\mathrm{s}=\{\mathrm{BNS,rCCSNe}\}$. Given initial guesses and other input parameters, the CCE model performs root finding to obtain the desired values of $M_{\mathrm{s}}$ that yield the target value of $F_{\rm BNS}(z(t),\theta)$ at $z=0$ and that guarantee the obtained abundance proxy to be consistent with the observed Milky-Way stellar abundances at $z = 0$, with $R_{\rm BNS,z0}$ as above (see Appendix~\ref{sec:GCE_evolution} for more details).

\begin{figure}[hbt!]
     \centering
     \includegraphics[width=0.47\textwidth]{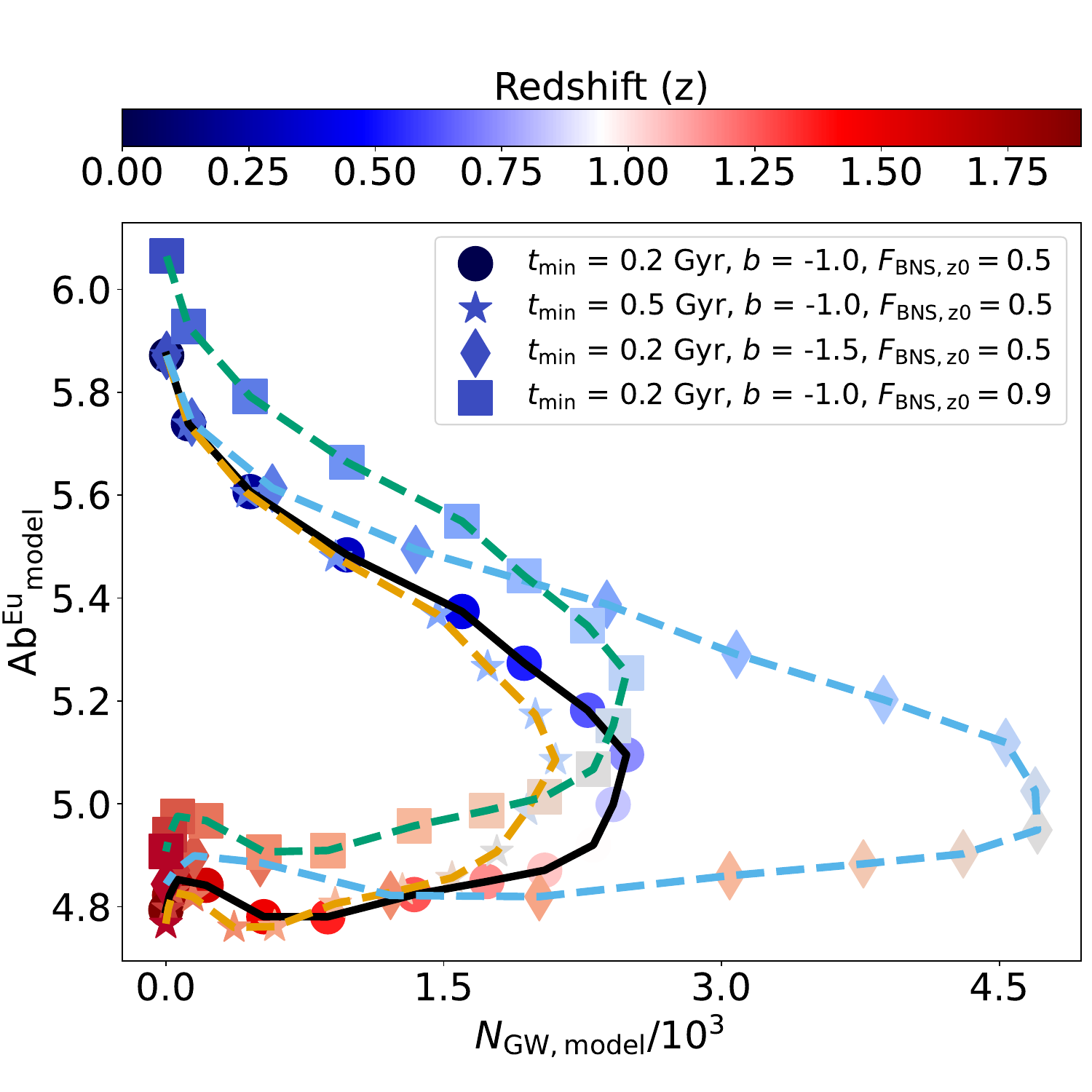}
     \includegraphics[width=0.47\textwidth]{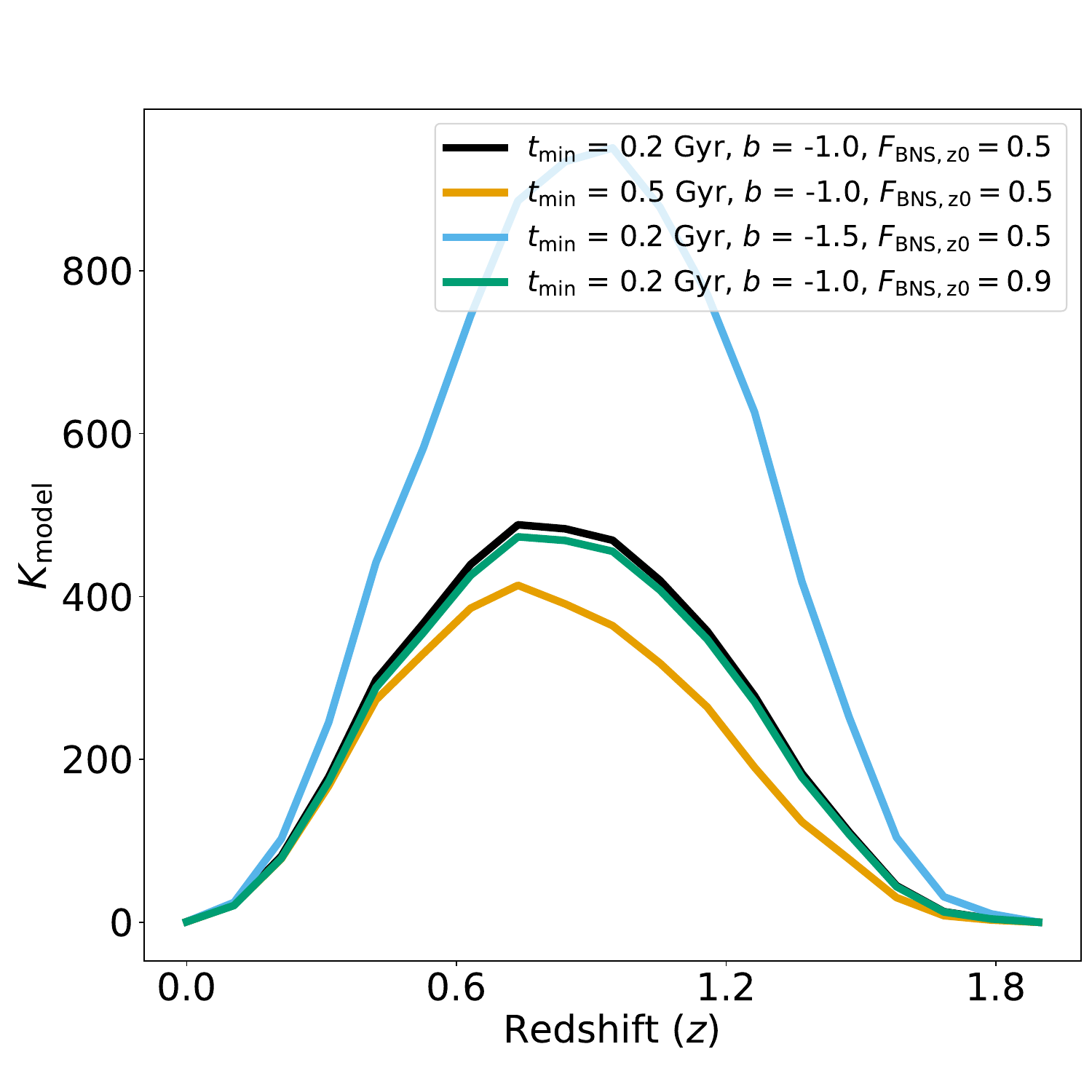}
     \caption{Modeled trajectories of total number of GW events ($N_{\mathrm{GW}}$ for 1 year of observations) and r-process abundance proxies ($\mathrm{Ab^{Eu}}$; top) as well as the corresponding modeled correlation functions $K_{\rm model}$ (bottom) as a function of redshift and for different DTD parameters $t_{\rm min}$, $b$ and fractional contributions $F_{\rm{BNS,z0}}$ of BNS to r-process enrichment. The colour bar on the top indicates the redshift of a particular point in the correlation phase space. The sensitivity of the trajectories in the $N_{\mathrm{GW}}$--$\mathrm{Ab^{Eu}}$ space to different parameters (particularly $F_{\rm{BNS,z0}}$) is evident, which forms the basis of inferring $F_{\rm{BNS,z0}}$ from observational data. }
     \label{fig:corr_demo_plot}
\end{figure}

The correlations $K_{\mathrm{model}}$ are then given by 
\begin{multline}
     K_{\mathrm{model}}(z;t_{\mathrm{min}},b,F_{\rm{BNS,z0}}) = \\
     \frac{N_{\rm GW,model}(z;t_{\mathrm{min}},b)}{{\mathrm{Ab}^{\mathrm{Eu}}}_{\mathrm{model}}(z;t_{\mathrm{min}},b,F_{\rm{BNS,z0}}) }.\label{eq:model_corr_full}
\end{multline}
In \figref{fig:corr_demo_plot}, the modeled $N_{\mathrm{GW,model}}$ and $\mathrm{Ab^{Eu}}_\mathrm{{model}}$ estimates are shown in the top panel for different values of the DTD parameters and $F_{\rm{BNS,z0}}$, with the corresponding correlation functions depicted in the bottom panel. 
The correlation functions peak at $z \approx 1$, corresponding to where the slope of the $N_{\mathrm{GW,model}}-\mathrm{Ab^{Eu}}_\mathrm{{model}}$ trajectories in the top panel undergo a sign change (or a `turning point'), due to a peak in the BNS merger rate. At smaller redshifts relative to this peak, the r-process contribution from BNS typically dominates over rCCSNe for the parameters considered here. Therefore, the position of the peak (or `turning point') in the $N_{\mathrm{GW,model}}$--$\mathrm{Ab^{Eu}}_\mathrm{{model}}$ space strongly constrains the contribution of BNS to the cosmic r-process. As the DTD parameters $b$ or $t_{\mathrm{min}}$ decrease for fixed $F_{\rm{BNS,z0}}$, the `turning point' remains almost constant in the $\mathrm{Ab^{Eu}}_\mathrm{{model}}$ direction, but the trajectories stretch in the $N_{\mathrm{GW,model}}$ direction due to enhanced BNS merger rates. The latter increase, since $R_{\rm BNS,z0}$ is fixed and the DTD steepens or extends to smaller minimum delay times. As a result, the peak of the correlation function also increases. 
For fixed DTD parameters, the overall BNS r-process fraction $F_{\rm{BNS,z0}}$ shifts the `turning point' of the trajectory in the $\mathrm{Ab^{Eu}}_\mathrm{{model}}$ direction. It is unchanged in the $N_{\mathrm{GW}}$ direction, since the BNS merger rate is not affected by $F_{\rm{BNS,z0}}$ in this case, with the average per-event yield of Eu being the only degree of freedom changing here. In particular, this illustrates the sensitivity of our method to the per-event r-process yield of BNS. 
The impact of $F_{\rm{BNS,z0}}$ on the trajectory slopes is qualitatively different from the impact of the DTD parameters and establishes the sensitivity of the correlation method to $F_{\rm{BNS,z0}}$. We discuss this point more quantitatively in \secref{sec:fisher Analysis}. 

The correlation trajectories $K_{\rm model}$ in \figref{fig:corr_demo_plot} illustrate the sensitivity to various parameters. The sensitivity of $K_{\rm model}$ to a change in $F_{\rm BNS,z0}$ appears to be smaller compared to its sensitivity to changes in the DTD parameters. However, this is only an effect of the $+5$ offset to the abundance proxy $\mathrm{Ab^{Eu}}$ added here. We verify in Appendix~\ref{app:coorelation_without_offset} that without the offset, a change in $F_{\rm BNS,z0}$ leads to a significant change in the correlation trajectory. Even though the correlation's sensitivity to variations in $F_{\rm BNS,z0}$ decreases due to the offset, the significance of the uncertainty estimates on $F_{\rm BNS,z0}$ from our proposed correlation model are much higher than those of the DTD parameters (see \secref{sec:fisher Analysis} and Table~\ref{tab:fisher_forecasts}). This is because the  $+5$ offset to $\mathrm{Ab^{Eu}}$ also leads to a decrease in the relative observational uncertainty of $\mathrm{Ab^{Eu}}$ compared to the original scale, which compensates for the reduced sensitivity due to the offset, as discussed in more detail in Appendix~\ref{app:coorelation_without_offset}.

\section{Parameter Uncertainty Estimation using Fisher Information}
\label{sec:fisher Analysis} 

We employ Fisher forecasts to determine the minimum uncertainty with which one can recover the maximum likelihood values of the parameters.\footnote{Whereas a Bayes estimate of posteriors on the parameters $\theta$ may be desirable when working with actual observational data in the future, here we choose to use the Fisher matrix to demonstrate the expected precision of the correlation method in the context of mock data.} According to the Cramer-Rao bound \citep{1945_Rao_bound_paper,1946_Cramer_bound_paper}, the minimum uncertainty in estimating the parameters around the maximum likelihood point ${\theta_{\rm ML}}$ is given by $\Delta \theta_p = \sqrt{(\mathds{F}^{-1})_{pp}}$,  where $\Delta \theta_p$ represents the standard deviation of our parameter estimates $\theta_{p}\in\{t_{\rm min},b,F_{\rm{BNS,z0}}\}$. For our scenario, the Fisher information matrix $\mathds{F}_{pq}$ is given by (see Appendix~\ref{app:Fisher_derivation})
\begin{align}
    \mathds{F}_{pq} =&  \bigg\langle\pdv{\log L}{\theta_p,\theta_q}\bigg\rangle_{\theta_{\rm ML}} \label{eq:derived_Fisher_matrix_form} \\
    =& \sum_{k = 1,2}\mskip-5mu 
     \sum_{z_n}  \ C_k^{-1}(z_n) \pdv{\mathcal{L}_{k}}{\theta_p}\pdv{\mathcal{L}_{k}}{\theta_q}\bigg|_{t_{\mathrm{min},0},b_0,F_{\mathrm{BNS,z0},0}}, \nonumber
\end{align}
where $\theta_p,\theta_q \in \{t_{\mathrm{min}},b,F_{\rm{BNS,z0}}$\} and where we have assumed that $\theta_{\rm ML}$ corresponds to the fiducial set of model parameters $\theta_{\mathrm{fiducial}}=\{t_{\mathrm{min},0},b_0,F_{\rm BNS,z0,0}\}$ described in \secref{subsec:Fisher_fiducial}. Here, $\mathcal{L}_{k}$ represent the models described in \secref{sec:GCE_modelling}, i.e., $(\mathcal{L}_{1},\mathcal{L}_{2}) \equiv (K_{\mathrm{model}} , \mathrm{Ab}^{\rm Eu}_{\mathrm{model}} )$. The covariance functions $C_k$ at redshift bin $z_n$ are given by Eqns.~\eqref{eq:covariance_matrix}--\eqref{eq:covariance_dprime_matrix}. They require mock estimates of future observations for $\hat{N}_{\rm GW}$ and $\mathrm{\hat{Ab}^{Eu}}$, which we discuss in \secref{sec:mock_estimates}, as well as of the associated measurement errors $\sigma_{\hat{N}_{\rm GW}}$, $\sigma_{\mathrm{\hat{Ab}^{Eu}}}$, and $\sigma_{d_{\rm L}}$, which we calculate in \secref{sec:expected_measurement_uncertainties}. We discuss parameter values in \secref{subsec:Fisher_fiducial}, which correspond to different possible astrophysical scenarios. Finally, the corresponding Fisher forecasts are summarized in \secref{subsec:Fisher_results}. 

\begin{table*}
	\centering
	\caption{The $1\sigma$-uncertainty on estimated parameters for fiducial DTD parameters $t_{\mathrm{min}}=0.2$\,Gyr, $b=-1$ and a combination of two CEs, one ET (2CE+ET), one year of observation time, and different assumed total fractional BNS contributions $F_{\rm{BNS,z0}}$ to the cosmic r-process. ``EM'' refers to the bright-siren case (presence of electromagnetic counterparts providing host galaxy redshift information). Numbers in parentheses represent the significance or signal-to-noise ratio (SNR) of the estimation.}
	\label{tab:fisher_forecasts}
	\begin{tabular}{lccr} 
		\hline
		Scenario
         & $\sqrt{\left(F^{-1}\right)_{F_{\rm{BNS,z0}}F_{\rm{BNS,z0}}}}$ (SNR) & $\sqrt{\left(F^{-1}\right)_{t_{\mathrm{min}}t_{\mathrm{min}}}}$ [Gyr] (SNR)& $\sqrt{\left(F^{-1}\right)_{bb}}$ (SNR)\\ 
        \hline
		 2CE + ET, $F_{\rm{BNS,z0}}=0.5$ & 0.03 ($2\times 10^1$) & 0.3 (0.7) & 0.2 ($0.5 \times 10^1$) \\ 

        2CE + ET, $F_{\rm{BNS,z0}}=0.9$ & 0.03 ($3\times 10^1$) & 0.3 (0.7) & 0.2 ($0.5 \times 10^1$)\\ 

        2CE + ET + EM, $F_{\rm{BNS,z0}}=0.5$ & 0.02 ($3\times 10^1$) & 0.2 (1) & 0.1 ($1 \times 10^1$)\\ 

        2CE + ET + EM, $F_{\rm{BNS,z0}}=0.9$ & 0.02 ($5\times 10^1$)& 0.2 (1) & 0.1 ($1 \times 10^1$)\\ 
       \hline
	\end{tabular}
\end{table*}

\subsection{Simulated Mock Estimates of Gravitational-Wave and Abundance Observations}
\label{sec:mock_estimates}

In the absence of observations of GW events and r-process abundances across a redshift range,  
we use the models ${N}_{\rm GW, model}$ and $\mathrm{{Ab}^{Eu}_{model}}$ to generate mock data. 
Individual future observations have associated observational errors that will shift the observed mean value within a redshift bin from the true or expected value. To include this effect in our mock estimates, we take $\mathcal{O}(10^3)$\footnote{The large sample size ensures that the random sampling error is converged, i.e., repeating the sampling process multiple times leads to similar statistical quantities each time.} samples from a normal distribution $\mathcal{N}(\mu(z_n),\sigma_{\mathcal{N}}^2(z_n))$ in each redshift bin $z_n$, taking the theoretically expected (modeled) mean values as $\mu$ and the expected measurement error as the standard deviation $\sigma_{\mathcal{N}}$. This generates $\mathcal{O}(10^3)$ Fisher matrix elements $\mathds{F}_{pq}$ (Eq.~\ref{eq:derived_Fisher_matrix_form}) and the corresponding mean value is used for deriving the 1$\sigma$ uncertainties (Cramer-Rao bound) on $F_{\rm BNS,z0},t_{\rm min}$, and $b$ in \secref{subsec:Fisher_results}. This sampling process generates mock estimates $\hat{N}_{\rm GW,mock}$ and $\rm{\hat{Ab}^{Eu}_{mock}}$ for $\hat{N}_{\rm GW}$ and $\mathrm{\hat{Ab}^{Eu}}$, respectively. To generate $\hat{N}_{\rm GW,mock}$ in a redshift bin $z_n$, the distribution $\mathcal{N}[\mu = N_{\rm GW, model}(z_n), \sigma^2 = \sigma^2_{\hat{N}_{\rm GW}}(z_n)]$ is sampled, where $\sigma^2_{\hat{N}_{\rm GW}}(z_n) = N_{\rm GW,model}(z_n)$ (\secref{subsec:data_covariance}). Similarly, the mock estimate $\mathrm{\hat{Ab}^{Eu}_{mock}} (z_n)$ of the abundance proxy is generated by sampling from the distribution $\mathcal{N}[\mu = \mathrm{Ab^{Eu}_{model}} (z_n),\sigma^2 = \sigma^2_{\mathrm{\hat{Ab}^{Eu}}}(z_n)]$, where $\sigma_{\mathrm{\hat{Ab}^{Eu}}}(z_n)$ is the expected uncertainty of the proxy. The abundance proxy represents the mean of the abundance observations of individual galaxies in the redshift bin $z_n$, hence $\sigma_{\mathrm{\hat{Ab}^{Eu}}}(z_n)$ is the standard error of the mean (SEM) of the abundance measurements in that redshift bin.

The SEM of the abundance proxy is affected by the intrinsic scatter of the abundance observations of individual galaxies in a given redshift bin $z_n$. The individual abundance observations of galaxies in a redshift bin will exhibit a standard deviation $\sigma_{\rm GAL}$ around the mean [$\mathrm{\hat{Ab}^{Eu}}(z)$] due to stochasticity of r-process events at low metallicity (rarity of events), inhomogeneous mixing of r-process material into the ISM, differences in star formation histories, galaxy masses, metallicity, etc. Assuming $\sigma_{\rm GAL}$ is typically much larger than the individual measurement uncertainty at all redshifts\footnote{The validity of this assumption is uncertain, but not implausible judging from abundance observations across metallicity and different environments in and around the Milky Way.}, the SEM is given by $\sigma_{\mathrm{\hat{Ab}^{Eu}}}(z_n) = \sigma_{\rm GAL}(z_n)/\sqrt{N_{\rm obs,Ab}(z_n)}$, where we assume $N_{\rm obs,Ab}(z_n)\ge 9$ number of observations in the redshift bin $z_n$.  
The abundance distribution in Milky Way disk stars with stellar metallicity above the solar value is [Eu/H] $\approx 0.2 \pm 0.06$ \citep{battistini_origin_2016,SAGA_database}. Assuming the mean abundance of such stars to be representative of the abundance proxy at $z=0$, this translates into $\sigma_{\rm GAL} (z=0) \approx 0.06$ and $\mathrm{\hat{Ab}^{Eu}_{model}}(z=0) \approx 5.2 $ 
for Milky-Way equivalent galaxies, with an expected SEM $\sigma_{\mathrm{\hat{Ab}^{Eu}}}(z=0)\lesssim 0.02$. We take the upper limit $\sigma_{\mathrm{\hat{Ab}^{Eu}}}(z=0) \approx 0.02$ for our mock estimate and obtain a relative SEM for Milky-Way-type galaxies of $\sigma_{\mathrm{\hat{Ab}^{Eu}}}(z=0)/{\mathrm{\hat{Ab}^{Eu}_{model}}}(z=0) \approx 0.004$. Due to the lack of abundance data for star-forming galaxies with $\tilde{\mathcal{M}}\gtrsim 8.5$ other than the Milky Way, we assume that the relative SEM of the Milky Way is representative of all galaxies at $z=0$. To model the $z>0$ values $\sigma_{\mathrm{\hat{Ab}^{Eu}}}(z)$, we assume the proxy is derived from electromagnetic (spectral) data. Since the measured spectral flux $F$ decreases with the source redshift as $F \propto (1+z)^{-2}$ and the Poisson error of photon counting statistics $\sigma_{F} \propto \sqrt{F}$ dominates detection errors, the relative error of the individual abundance measurements scales as $(\sigma_{F}/F) \propto (1+z)$ \footnote{The scaling is also motivated by the $(1+z)$ scaling of uncertainties in spectral quantities such as the photometric redshift, assumed in spectroscopic surveys such as \cite{2006A&A_Ilbert_photo_z_errors,2021A&A_miniJPAS_survey}.}. The standard deviation $\sigma_{\rm GAL}$\footnote{$\sigma(\mathds{a}\mathcal{X}) = \mathds{a}\sigma(\mathcal{X})$, where $\mathcal{X}$ is a random variable and $\mathds{a}$ is a constant factor (here, $\mathds{a}=1+z$, which is constant in each redshift bin). However, the rarity of r-process events at low metallicity and the difficulty of measuring abundances especially at significant redshifts $z\gtrsim 0.1$ might, due to low number statistics, lead to a noisy standard deviation, that is, a larger value of $\sigma_{\rm GAL}$.} 
and the relative SEM also follow the same scaling with redshift. Therefore, we obtain the redshift-dependent uncertainty $\sigma_{\mathrm{\hat{Ab}^{Eu}}} (z) \approx 0.004 (1+z)\mathrm{{Ab^{Eu}}_{model}}(z)$.

\subsection{The Covariance Functions \texorpdfstring{$C_k$}{Ck}}
\label{sec:expected_measurement_uncertainties}

The observational uncertainties $\sigma^2_{\hat{N}_{\rm GW}}(z_n)$ and $\sigma_{\mathrm{\hat{Ab}^{Eu}}} (z_n)$ as modelled in \secref{sec:mock_estimates} contribute to the covariance functions $C_k$ (Eqns.~\eqref{eq:covariance_matrix}--\eqref{eq:covariance_dprime_matrix}) through Eqns.~\eqref{eq:grav_sources_errors} and  \eqref{eq:abundance_error}. The remaining, unmodelled uncertainty of the luminosity distance measurements ($\sigma_{d_{\rm L}}$, \equnref{eq:redshift_error}) is calculated from a mock GW database as described in Appendix~\ref{sec:Mock_GW_catalogue}. For this mock database, we assume a detector combination of two CE observatories and one ET in generating data for the number of GW events, $\hat{N}_{\mathrm{GW,mock}}$. The detector configuration affects the GW detector selection function $\phi(z)$ as calculated in Appendix~\ref{sec:Mock_GW_catalogue}.

Examining the covariance functions $C_k$ overall, the assumed 10\% relative uncertainty in the mean r-process abundance ([Eu/H] $= 0.2\pm 0.02$) at $z = 0$ and its scaling $\propto(1+z)$, makes it the dominant observational uncertainty in this analysis. Hence, an increase or decrease in the mean r-process abundance proxy errors will have a proportional effect on the overall precision of parameter estimation with this method.

\subsection{Fiducial Parameters for Fisher Forecast} \label{subsec:Fisher_fiducial}

In addition to the models $\mathcal{L}_k$, the covariance functions $C_k$, and the mock estimates of $\hat{N}_{\rm GW}$ and $\mathrm{\hat{Ab}^{Eu}}$, the Fisher matrix elements (\equnref{eq:derived_Fisher_matrix_form}) require a fiducial set of parameter choices. As a fiducial astrophysical scenario, we choose the model parameters $\theta_{\mathrm{fiducial}}$ with $t_{\mathrm{min},0} = 0.2$\,Gyr, $b_0 = -1$, and $F_{\rm BNS,z0,0} = 0.5$. The fiducial assumption of $t_{\mathrm{min}} = 0.2$\,Gyr is consistent with $t_{\mathrm{min}} = 184^{+67}_{-79}$\,Myr determined by \citet{Zevin_dtd_sgrb_2022} based on an analysis of short GRB host galaxy associations. The fiducial value $b = -1$ is incompatible with the results of \citet{Zevin_dtd_sgrb_2022}. However, it represents the standard expectation (see \secref{sec:GCE_modelling}) and it is within the range of the parameter bounds determined by other short GRB and DTD studies (e.g., \citealt{2015_Wanderman_Piran_sGRB_pop_DTD,2020_paterson_DTD_GRB_afterglow,2019_Beniamini_Piran_BNS_DTD, 2019_Safarzadeh_Paper_III, 2020_McCarthy_DTD_from_host_galaxy_properties}). Although BNS mergers likely contribute an order unity fraction to the Galactic r-process, a dominant contribution is incompatible both with spectroscopic observations of high-metallicity stars in the Galactic disk as well as with low-metallicity stars in the Galactic halo (\citealt{wehmeyer_galactic_2015,cote_neutron_2019,2019Nature_Siegel,vandevoort_neutron_2020,yamazaki_possibility_2022,brauer_collapsar_2021,chen_inference_2025}; though cf.~\citealt{shen_history_2015-1,duggan_neutron_2018,macias_constraining_2019,bartos_early_2019,tarumi_evidence_2021}). Even fast merging populations of BNS systems (e.g., \citealt{2019_Beniamini_Piran_BNS_DTD,2025_Maoz_Ehud_DTD_paper}) are inconsistent with current GW observations and abundance observations of the Galactic disk \citep{saleem_mergers_2025}. We therefore choose a moderate fiducial value of $F_{\rm{BNS,z0}}=0.5$ here. The effects of other choices for the value of $F_{\rm{BNS,z0}}$ on our analysis are discussed in Appendix~\ref{app:sensitivity}. We also investigate the effect of varying our fiducial assumptions regarding the BNS DTD parameters on the precision of our Fisher estimates in Appendix~\ref{app:sensitivity}.

\begin{figure}
    \centering
    \includegraphics[width = 0.49\textwidth]{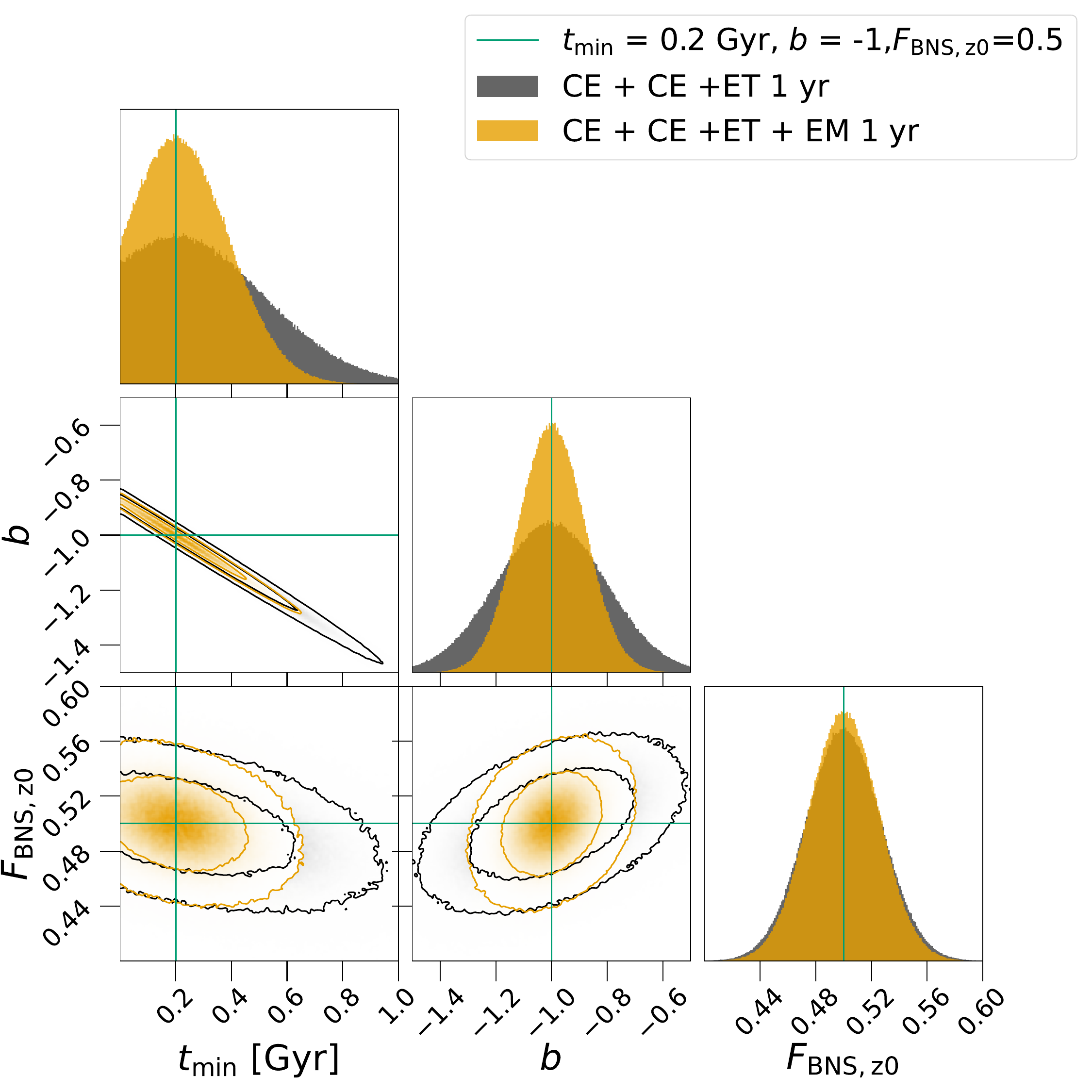}
    \caption{Fisher forecasts demonstrating the precision with which the proposed method can estimate the values of the BNS DTD parameters $t_{\mathrm{min}}$ and $b$ as well as the total fraction $F_{\rm{BNS,z0}}$ that BNS contribute to cosmic r-process enrichment. The forecasts assume one year of observation time and two CE observatories as well as one ET observatory. Results are shown for cases with and without EM counterparts. The panel for $t_{\mathrm{min}}$ has been truncated to display only positive (meaningful) parameter values. Contour lines represent the $1 \sigma$ and $2 \sigma$ uncertainty ranges. Green lines represent the fiducial values of the parameters.}
    \label{fig:Fisher Analysis}
\end{figure}

\subsection{Uncertainty of Parameter Estimation from Fisher Information Matrix}
\label{subsec:Fisher_results}

 The 1$\sigma$-uncertainties (Cramer-Rao bound) on the parameters for our fiducial choice $t_{\mathrm{min}} = 0.2$\,Gyr, $b=-1$, and $F_{\rm{BNS,z0}}=0$ in scenarios with (`bright sirens') and without (`dark sirens') EM counterparts, are presented in Table~\ref{tab:fisher_forecasts}. We also present the more `conservative' choice of $F_{\rm{BNS,z0}}=0.9$, which increases the correlation $K_{\rm model}$ between the redshift-dependent number of BNS GW events and cosmic r-process enrichment, thus enhancing the sensitivity (increased signal-to-noise, SNR) of our method to $F_{\rm{BNS,z0}}$. More details on varying the fiducial parameters (astrophysical scenarios) are discussed in Appendix~\ref{app:sensitivity}. 
 
 Table~\ref{tab:fisher_forecasts} shows that our method is capable of constraining the total cosmic BNS r-process contribution to the $\lesssim 6\%$ level, and generally to better than the $\lesssim\!10\%$ level (Table~\ref{tab:fisher_sensitivity}, Appendix~\ref{app:sensitivity}). 
 Regarding the DTD parameters, our Fisher forecasts for the power law index $b$ are comparable to the parameter estimation bounds from previous studies based on GRBs (e.g. \citealt{Zevin_dtd_sgrb_2022, 2015_Wanderman_Piran_sGRB_pop_DTD,2019_Safarzadeh_paper_II,2020_McCarthy_DTD_from_host_galaxy_properties}), whereas that for the minimum delay time $t_{\rm min}$ and its SNR are worse by at least a factor of a few. This is because of the average delay $\langle t_{\rm delay} \rangle$ of BNS mergers, 
 \begin{align}
     \langle t_{\rm delay} \rangle &= \frac{\int_{t_{\rm min}}^{t_{\rm max}} t' \  D(t',t_{\rm min},b) \ \mathrm{d}t'}{\int_{t_{\rm min}}^{t_{\rm max}} D(t',t_{\rm min},b) \ \mathrm{d}t'} \nonumber\\
     &=\begin{cases}
     \frac{b+1}{b+2}\left(\frac{t_{\rm max}^{b+2}-t_{\rm min}^{b+2}}{t_{\rm max}^{b+1}-t_{\rm min}^{b+1}}\right) \ & \mbox{for }\ b \neq-1,-2 \\
     \frac{t_{\rm max}-t_{\rm min}}{\ln t_{\rm max}-\ln t_{\rm min}} \ & \mbox{for }\ b = -1 \\
     \frac{\ln t_{\rm max}-\ln t_{\rm min}}{t^{-1}_{\rm min}-t_{\rm max}^{-1}} \ & \mbox{for }\ b = -2
     \end{cases},\label{eq:av_delay}
 \end{align}
 where $t_{\rm max} = t(z=0)$ is the cosmic age at $z=0$, is largely insensitive to $t_{\rm min}$ relative to $b$ when $|b|\gtrsim 1$ and $t_{\rm min}\ll t_{\rm max}$.  The fractional sensitivity of this average delay to variations in $t_{\rm min}$, $\frac{\partial  \langle t_{\rm delay} \rangle}{\partial t_{\rm min}}\frac{t_{\rm min}}{ \langle t_{\rm delay} \rangle}=\frac{\partial \ln  \langle t_{\rm delay} \rangle}{\partial \ln t_{\rm min}}$, is insignificant compared to $\frac{\partial \ln \langle t_{\rm delay} \rangle}{\partial \ln |b|}$ for the fiducial parameter values ($t_{\rm min}\ll t_{\rm max}$). This results in reduced sensitivity of our correlation method to $t_{\rm min}$ and thus in low significance Fisher forecasts for $t_{\rm min}$ in Table ~\ref{tab:fisher_forecasts}. However, as the value of $t_{\rm min}$ increases, the fractional sensitivity to variations in $t_{\rm min}$ increases; at a parameter choice of $t_{\rm min} = 1.0 \ $ Gyr, all other parameters remaining at their fiducial values, the relative uncertainty of $t_{\rm min}$ drops to $40\%$ as reported in Table~\ref{tab:fisher_sensitivity}.

 \begin{figure}
    \centering
    \includegraphics[width = 0.5\textwidth]{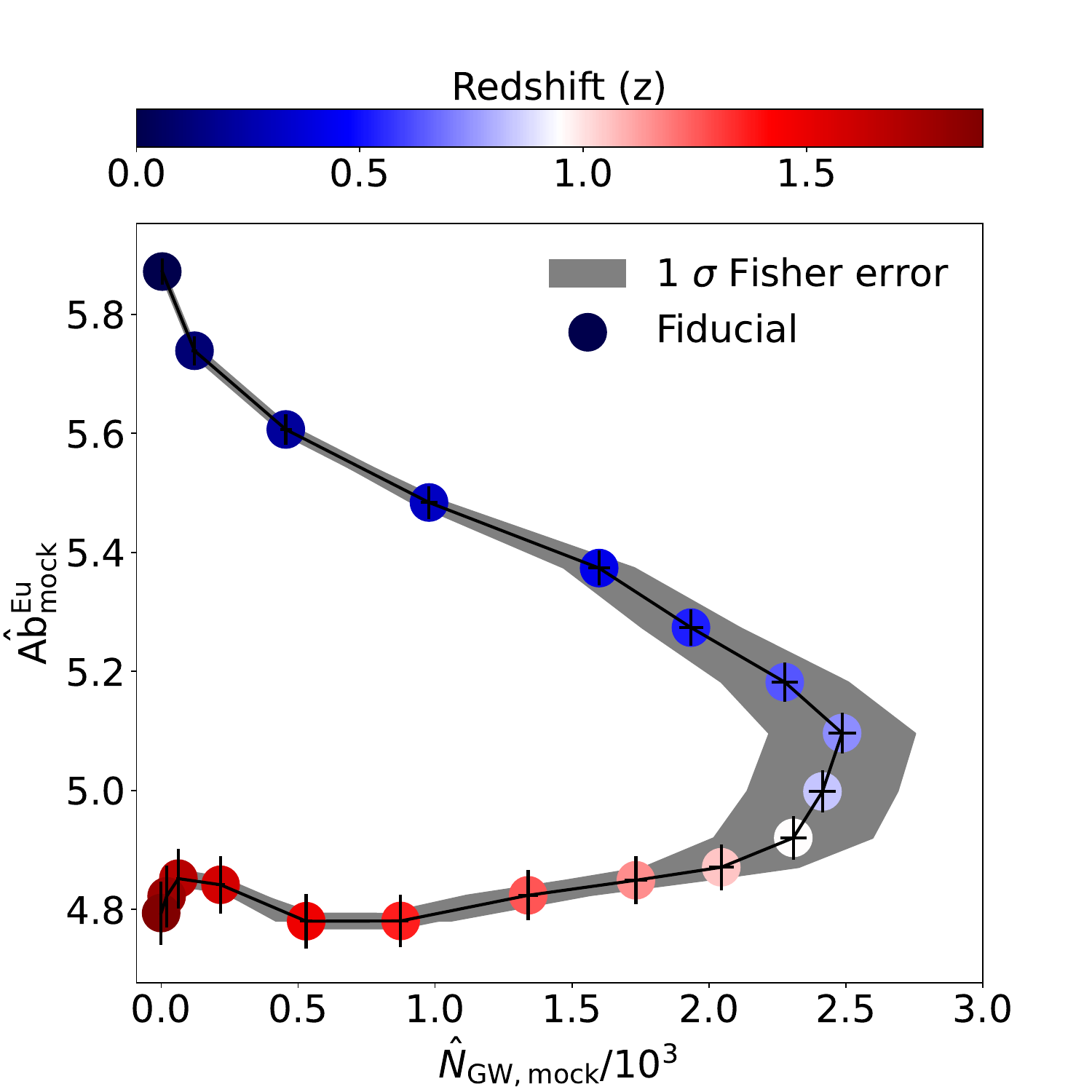}
    \caption{Fiducial mock data correlation trajectory (parameters $t_{\rm min}=0.2$\,Gyr, $b=-1$, $F_{\rm BNS,z0}=0.5$) assuming two CE detectors, one ET, one year of observation time, and no EM counterparts. Error bars represent the uncertainties as discussed in Secs.~\ref{subsec:data_covariance} and \ref{sec:mock_estimates}. The gray-shaded band shows the $1\sigma$ uncertainty range of the estimated trajectory as calculated by sampling from the joint distribution obtained from the Fisher matrix analysis of the estimated parameters (first row of Table~\ref{tab:fisher_forecasts}; Fig.~\ref{fig:Fisher Analysis}).} 
    \label{fig:K_prof}
\end{figure}

 \begin{figure}
     \centering
     \includegraphics[width = 0.5\textwidth]{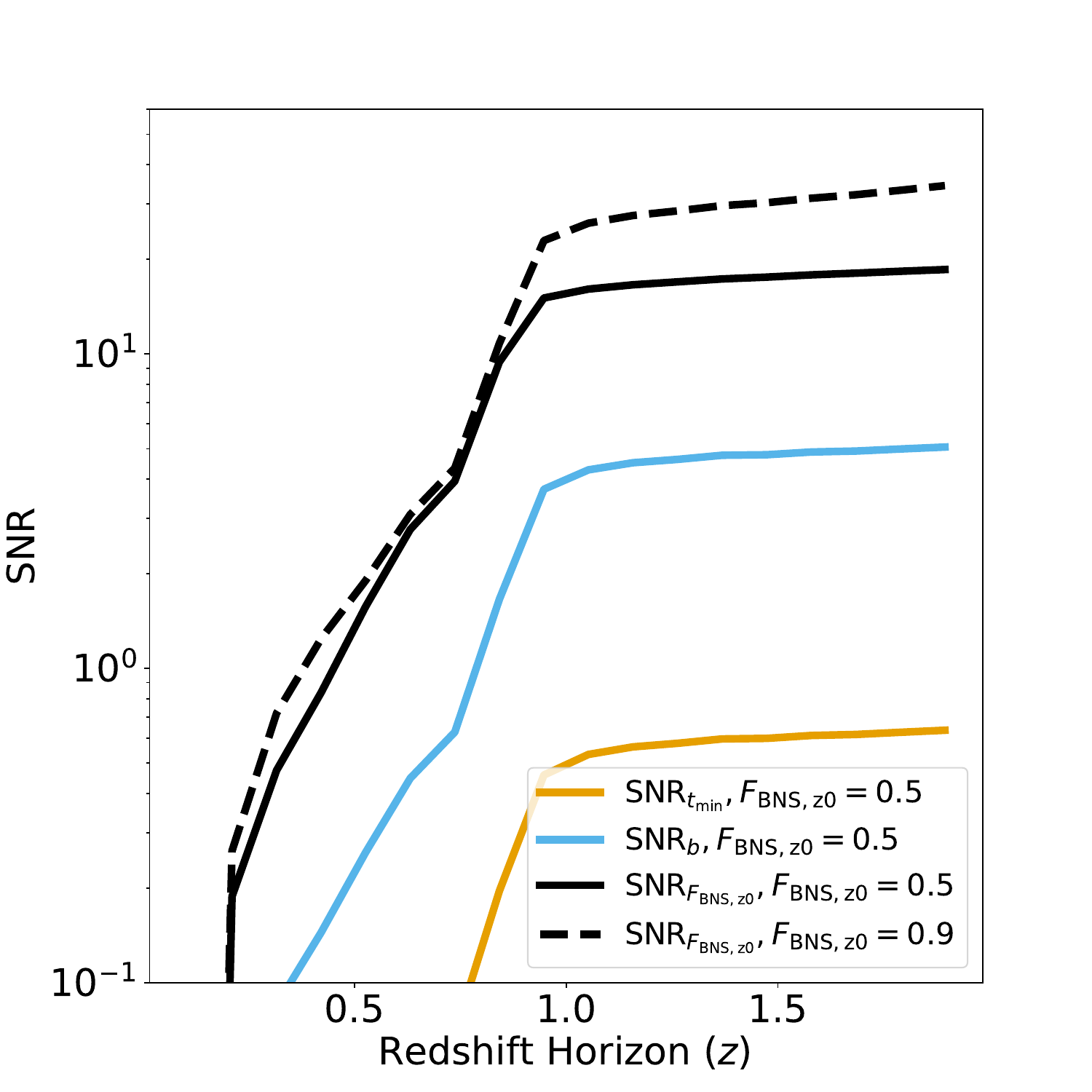}
     \caption{The expected signal-to-noise ratio (SNR) for inferring $F_{\rm{BNS,z0}}$, $t_{\rm min}$ and $b$ with our correlation method as a function of the redshift horizon of future r-process and GW observations and for a combination of two CEs and one ET (2CE+ET), adopting the fiducial astrophysical scenario $F_{\rm{BNS,z0}} = 0.5$, $t_{\mathrm{min}}=0.2$, and $b=-1$ with one year of observation time. For comparison, the SNR for inferring $F_{\rm{BNS,z0}}$ from an almost exclusively BNS dominated r-process scenario ($F_{\rm{BNS,z0}} = 0.9$; otherwise identical to the fiducial one) is also shown.}
     \label{fig:SNR_max_redshift}
 \end{figure}

 Figure \ref{fig:Fisher Analysis} shows the Cramer-Rao bound on the parameter estimation for a combination of two CE detectors and one ET observatory assuming 1 year of observation time. The DTD parameters can be estimated with higher precision in the bright-siren case (EM counterparts providing host galaxy redshifts) due to reduced redshift uncertainty of the GW observations. The parameter bounds on $F_{\rm{BNS,z0}}$ are almost identical with or without EM signals, because the estimation precision of $F_{\rm{BNS,z0}}$ (determined by the correlation $\mathcal{L}_1$ and the abundance term $\mathcal{L}_2$) is dominated by the conservative uncertainty estimate of the r-process abundance proxy. Hence, the reduced redshift uncertainties from the EM signal do not play a significant role for the precision in estimating $F_{\rm{BNS,z0}}$.

Figure~\ref{fig:Fisher Analysis} shows a strong negative parametric dependence correlation between the DTD parameters. A decrease in $b$ is balanced by an increase in $t_{\mathrm{min}}$ to accommodate the mock data and vice versa. The absence of a strong parametric dependence correlation between $F_{\rm{BNS,z0}}$ and the DTD parameters is due to the fact that $F_{\rm{BNS,z0}}$ is controlled by the volumetric rate--yield $M_{\mathrm{s}}$ of BNS and rCCSNe, $\mathrm{s}=\mathrm{\{BNS,rCCSNe\}}$ (see Eq.~\eqref{eq:M_s}).

Figure \ref{fig:K_prof} shows the correlation trajectory of our fiducial mock data with modelled observational error bars as described in Secs.~\ref{subsec:data_covariance} and \ref{sec:mock_estimates}. Overlaid as a gray-shaded region is the 1$\sigma$-uncertainty range in recovering the trajectory from Fisher parameter estimation. It corresponds to the range of model trajectories obtained by sampling parameters within the Fisher bounds as reported in the first row of Tab.~\ref{tab:fisher_forecasts}.
These results illustrate the sensitivity of our method to different parameter values in the $1\sigma$ Fisher forecast range. As discussed in \secref{sec:GCE_modelling}, a change in DTD parameters causes a trajectory to shift in the $\hat{N}_{\mathrm{GW,mock}}$ direction, whereas a variation in $F_{\rm{BNS,z0}}$ has an approximately orthogonal effect. Since our Fisher forecasts predict smaller relative errors in the latter parameter, the gray region in \figref{fig:K_prof} has a relatively narrow width in the $\mathrm{\hat{Ab}^{Eu}_{mock}}$ direction relative to the $\hat{N}_{\mathrm{GW,mock}}$ direction; the DTD parameters are weakly constrained relative to $F_{\rm BNS,z0,0}$, a general property that is evident from the SNR values in Tab.~\ref{tab:fisher_forecasts}.

We conclude this section by briefly elaborating on the requirements for future observations to constrain the BNS contribution $F_{\rm BNS,z0}$. Figure \ref{fig:SNR_max_redshift} reports the redshift-dependent SNR for inferring $F_{\rm{BNS,z0}}$, $t_{\rm min}$ and $b$ across a wide range of astrophysical scenarios (fractional contributions of BNS to the cosmic r-process). We find that largely irrespective of the astrophysical scenario, observations of r-process abundances and GWs out to $z\lesssim 0.7$ are required to provide an estimate on $F_{\rm{BNS,z0}}$ with $\lesssim 20\%$ relative error at $1\sigma$. Similarly,  observations out to $z\lesssim 1.0$ can provide an estimate on $b$ with $\lesssim 20\%$ relative error and an absolute error of $\Delta t_{\rm min} = 0.4 \ $Gyr on $t_{\rm min}$ ($\approx 200 \%$ relative error) at $1\sigma$. The $t_{\rm min}$ parameter estimation uncertainty is large because, as discussed earlier, our correlation method is only weakly sensitive to $t_{\rm min}$ at the fiducial parameter value of 0.2\,Gyr. 

\section{Discussion and Conclusion}
\label{sec:conclusion}

Whereas previous work has focused on assessing the contribution of BNS mergers to r-process nucleosynthesis in various local environments, such as stars in the Milky Way disk (e.g., \citealt{cote_origin_2018,hotokezaka_neutron_2018,2019Nature_Siegel,siegel_gw170817_2019,cote_neutron_2019,2021_Wanajo_NS_mergers_r_process,2021_Dvorkin_impact_of_turbulent_mixing,kobayashi_can_2023,2024ApJ_Holmbeck_total_r_process_yields_BNS,chen_inference_2025,saleem_mergers_2025}), metal-poor stars of the halo (e.g., \citealt{ishimaru_neutron_2015,tarumi_evidence_2021,2022ApJ_Amend_r_process_rain_from_BNS}), and dwarf galaxies (e.g., \citealt{2014_Tsujimoto_r_process_enrichment_paper,naidu_evidence_2022}),  here we propose a novel method to quantify the BNS contribution to the \emph{cosmic} r-process across cosmic history.

Our method exploits the fact that BNS mergers as sources of r-process elements and GWs introduce a non-trivial correlation between r-process abundances of galaxies and the population of GW events as a function of redshift. These correlations are expected to be largely independent of the underlying (cosmic) star formation history (Sec.~\ref{sec:correlation_function_definition}). Rather, they strongly depend on the overall contribution of BNS to the total r-process enrichment in the Universe and on the DTD of BNS mergers relative to star formation. We devise a cosmic chemical evolution model (Appendix~\ref{sec:GCE_evolution}) and model the redshift-dependent correlations $K(z)$ as a function of the cumulative fractional contribution $F_{\rm{BNS,z0}}$ of BNS to the cosmic r-process up to $z=0$ and of the BNS DTD parameters $t_{\rm min}$ and $b$ (\secref{sec:GCE_modelling}). The complimentary contribution to the r-process is assumed to be generated by a star-formation tracking (`prompt') source associated with the death of massive stars, such as collapsars or magnetorotational supernovae (collectively termed rCCSNe here). Empirically, the presence of multiple sources with different properties regarding their behaviour relative to star formation (`delayed' vs.~`prompt') is not solely motivated by results of chemical evolution analyses (see above mentioned references), but also by the decreasing inferred local BNS merger rate due to non-detections of confirmed BNS mergers through GWs \citep{2025_GWTC4_r_and_p}. Given current rates, the contribution of non-BNS, `prompt' sources is likely required even to explain the observed total current r-process abundances in Galactic stars (\citealt{chen_inference_2025,saleem_mergers_2025}; however, see also \citealt{2026arXiv_zenati_sub_dominant_r_process}). Given future observational data at $z\gtrsim 0.1$, models of $K(z)$ can be fit to infer the BNS DTD and the value of $F_{\rm{BNS,z0}}$, thus quantifying the role of BNS in cosmic r-process nucleosynthesis. 

The correlation method presented here is designed in view of the 3G GW detector era. For the time being, we construct mock estimates of the desired data and use Fisher forecasts to predict the expected uncertainty in estimating $F_{\rm{BNS,z0}}$ and the DTD parameters. We assume a combination of two CEs and one ET over one year of observation time. As our best-case scenario, we assume that all GW events are accompanied with an EM counterpart (e.g., afterglow of a short GRB, orphan afterglow, kilonova) that allows for a direct inference of the redshift (`bright-siren' case). Exploiting the full redshift range of the correlation, we obtain a relative uncertainty range of $\delta F_{\rm{BNS,z0}}/F_{\rm{BNS,z0}} \leq 5\%$ at $1\sigma$ and of $\approx 100 \%$ on $t_{\rm min}$ and $\approx 10 \%$ on $b$ at $1\sigma$, for a fiducial astrophysical scenario of $F_{\rm{BNS,z0}}=0.5$, $t_{\rm min}=0.2$\,Gyr, and $b = -1$ (Sec.~\ref{subsec:Fisher_results}; Tab.~\ref{tab:fisher_forecasts}; Fig.~\ref{fig:Fisher Analysis}). In the more pessimistic `dark-siren' scenario (absence of EM counterparts and direct redshift measurements), we instead obtain $\delta F_{\rm{BNS,z0}}/F_{\rm{BNS,z0}} \leq 6\%$ at $1\sigma$ and of $\approx\! 150 \%$ on $t_{\rm min}$ and $\approx\! 20 \%$ on $b$ at $1\sigma$. These results are largely independent of the assumed astrophysical scenario, except that the significance of inferring $F_{\rm{BNS,z0}}$ increases strongly with increasing BNS contributions $F_{\rm{BNS,z0}}$ (Appendix \ref{app:sensitivity}).

One limitation of our current Fisher information approach is the assumption of Gaussianity of the likelihood. Whereas this assumption may break down for actual future abundance data at increasing redshift due to availability of observations, intrinsic scatter owing to the rarity of enrichment events, and large error bars of individual measurements, it will at least be valid for GW data according to the central limit theorem, given the expected $\mathcal{O}(10^3)$ BNS events detected by 3G detectors. Future work with this method will employ Bayesian inference using Monte-Carlo methods, which can account for non-Gaussian effects on the likelihood. Such inference will also be able to consistently measure all parameters involved, including the per-event yield of BNS and non-BNS r-process sources.

Our method is more general than the somewhat specific illustration presented here; it allows for several extensions, and has astrophysical implications beyond solely quantifying the BNS r-process contribution. For instance, whereas here we constrain the fractional contribution of BNS to the r-process at $z=0$, one can simultaneously quantify that fraction over cosmic time as a function of redshift $F_{\rm BNS}(z)$, as well as the contribution $F_{\rm rCCSNe}(z)$ of `prompt' sources. Furthermore, the method can be generalized to multiple r-process proxies and correlations among them can be explored over cosmic time. These could include observational signatures that trace groups of elements (e.g., light vs.~heavy r-process elements), or even individual elements. Furthermore, it is straightforward to include more astrophysical channels of r-process production (e.g.~NS--BH mergers and consider subclasses of CCSNe) and different DTDs, and infer their support by the data. 

The inference of BNS DTD parameters as enabled by our method has broader consequences. They are tightly related to the properties of their host galaxies \citep{2020_Adhikari_dtd_papers,2020_McCarthy_DTD_from_host_galaxy_properties,2020MNRAS_Artale_host_galaxy_props,2022_broekgaarden_DCO_source_props_and_GW}. The uncertainty in BNS DTD parameters leads to a spread in the mass function of BNS host galaxies  \citep{2021_DTD_and_host_galaxy_mass_function}, and a reduction in this uncertainty can help to better identify BNS host galaxies and perform targeted follow-ups of future BNS-based multi-messenger events. The capability of constraining the minimum delay time to $~\mathcal{O}(100)$\,Myr as demonstrated here can provide information on a potential rapidly merging subpopulation of BNS (`rapid mergers'; e.g., \citealt{2019_Beniamini_Piran_BNS_DTD,2025_Maoz_Ehud_DTD_paper}). Accurate knowledge of the redshift distribution of BNS mergers (as a result of accurate DTD estimation) can help constrain cosmological parameters with standard siren cosmology, as pointed out by \cite{2019_Ding_standard_siren_with_redshift_dist}. The study of BNS formation channels, including the presence of common-envelope phases, are affected by assumptions of the BNS DTD \citep{2012_Dominik_DCO_I}, and hence will be informed from reduced uncertainties in the delay time parameters. 

Overall, the inference results discussed here sensitively depend on the availability of future r-process abundance data and their observational uncertainties (which we model here with a conservative scaling $\propto 1+z$ based on local r-process measurements). Our sensitivity analysis of parameter inference to the redshift horizon of joint GW and abundance observations (\secref{subsec:Fisher_results}; Fig.~\ref{fig:SNR_max_redshift}) demonstrates that, largely irrespective of the astrophysical scenario, the SNR is mainly acquired using data within $z \lesssim 1.0$. More specifically, observations out to $z\lesssim 0.7$ are required to constrain $F_{\rm BNS,z0}$ to better than $\lesssim 20\%$ at $1\sigma$. Observations to $z \lesssim 1.0$ can yield $t_{\rm min}$ and $b$ to about $\approx 200\%$ and $\lesssim 20\%$ at $1\sigma$, respectively. The 3G GW detectors will be able to detect BNS mergers out to redshifts $\lesssim 2$ \citep{2019Sathyaprakash_cosmology_in_early_universe,2019_Cosmic_explorer_white_paper,2020_Einstein_Telescope_white_paper,2021arXiv_Evans_horizon_study, 2022_3G_detection_capabilities}, but whether r-process abundance measurements outside the local group and out to redshifts $z \lesssim 0.7$ will materialize in the near future remains an open question. Progress on measuring abundances of $\alpha$-elements and metallicity using galaxy continuum emission \citep{1998ApJ_Heckman,2004ApJ_rix_spectral_modeling, 2006MNRAS_Crowther_on_the_reliability, 2012A&A_Sommariva_stellar_metallicity, 2019MNRAS_cullen_vandels}  as well as from the gas phase of emission-line galaxies out to redshifts $\lesssim 2$ and beyond \citep{2023ApJ_Isobe_JWST_identification,2024A&A_Schaerer_N_emitter,2024MNRAS_JI_GA-NIFS,2024A&A_Marques-chaves_N_emitters,2024A&A_Velichko_alpha_element_abundances,2025A&A_Curti_JADES_JWST_SFR} as well as exploiting correlations between elemental abundances using indirect approaches, e.g., also via cosmological simulations, may not render this challenge impossible. At the very least, our method could be applied to abundance evolution reconstructed with stellar archaeology from local observations of metal-poor stars, but it motivates identifying direct or indirect signatures of neutron-capture elements in the emission of galaxies beyond the local Universe.

{\it Code Used:} \texttt{Planck18} cosmology module in \texttt{astropy} \citep{astropy:2013, astropy:2018, astropy:2022}, \textit{SciPy: }\cite{2020SciPy-NMeth}. 

{\it Acknowledgements:} The authors gratefully acknowledge the computing time made available to them on the high-performance computers ``Emmy'' and ``Lise'' at the NHR Centers NHR@Göttingen and NHR@ZIB. These Centers are jointly supported by the Federal Ministry of Education and Research and the state governments
participating in the National High-Performance Computing (NHR) joint funding program (http://www.nhr-verein.de/en/our-partners). 
The work of S. M. is a part of the $\boldsymbol{\langle}\texttt{data|theory}\boldsymbol{\rangle}$ \texttt{Universe-Lab} which is supported by the TIFR and the Department of Atomic Energy, Government of India. We acknowledge the support of the Department of Atomic Energy, Government of India, under Project Identification No. RTI 4012. This research is supported by the Prime Minister Early Career Research Award, Anusandhan National Research Foundation, Government of India.  

\section*{Data Availability}

The data underlying this article are available in this article. The observational data is presented with relevant citations and is publicly available.


\bibliographystyle{mnras}
\bibliography{references,gr-astro}




\appendix

\section{Cosmic chemical evolution model}
\label{sec:GCE_evolution}

We model the cosmic r-process abundance proxy by generalizing the one-zone galactic chemical evolution model described in \citet{2019Nature_Siegel} to the cosmic context of galaxy distributions. Europium masses and number densities are first modelled as a function of cosmic age $t$ in a galaxy of arbitrary stellar mass, including BNS and rCCSNe as sources. By taking the galaxy density $N(\mathcal{\tilde{M}},t)$ (number of galaxies per logarithmic galaxy stellar mass $\mathcal{\tilde{M}} = \log_{10}(M/M_{\odot})$ per comoving volume $V_c$) and integrating over galaxy mass and comoving volume, we obtain a cosmic r-process abundance proxy as a function of time (or redshift). In generalization of \citet{1974PS_function}, the galaxy stellar mass function provides the galaxy density

\begin{align}
    N(\mathcal{\tilde{M}}, t) &= \frac{\mathrm{d}^2n}{\mathrm{d}\mathcal{\tilde{M}}\mathrm{d}V_c} \label{eq:Press_Schechter} = \begin{cases}
        \zeta_1  & \mbox{for $z(t) \le  3.5$} \\
        \zeta_2 & \mbox{for $z(t)> 3.5$} \\
    \end{cases},
\end{align}
where
\begin{align}
    \zeta_1 &= \ln(10)\exp\Big[-10^{(\mathcal{\tilde{M}}-\mathcal{M}^{\ast}(t))}\Big] \\
        &\times \Big \{ \,N _1^*\,\Big[10^{(\mathcal{\tilde{M}} - \mathcal{M}^{\ast}(t))}\Big]^{\alpha _1(t)+1} \nonumber + N _2^*\,\Big[10^{(\mathcal{\tilde{M}} - \mathcal{M}^{\ast}(t))}\Big]^{\alpha _2(t)+1}\Big\} ,
\end{align}
and
\begin{align}
    \zeta_2 &=&  \ln(10)\exp\Big[-10^{(\mathcal{\tilde{M}}-\mathcal{M}^{\ast}(t))}\Big]
        \times N^{\ast}(t) \Big[10^{(\mathcal{\tilde{M}} - \mathcal{M}^{\ast}(t))}\Big]^{\alpha(t)+1} ,\nonumber\\
\end{align}
as adapted from Eq.~(7)--(8) of \citet{2023A&A_COSMOS_GSMF_paper}. Here, $n$ is the number of galaxies, and the logarithmic cutoff mass $\mathcal{M}^{\ast}(t)$ below which astrophysical structures are expected to form has the same units as $\mathcal{M}$. The normalizations $N^{\ast}(t)$, $N_1^{\ast}(t)$, and $N_2^{\ast}(t)$ in $\rm{dex^{-1}Mpc^{-3}}$ and the exponents $\alpha(t)$, $\alpha_1(t)$, and $\alpha_2(t)$ are taken from the COSMOS2020 survey results for redshifts bins $z_n<5.5$ (Table C.2 of \citealt{2023A&A_COSMOS_GSMF_paper}) and from Table C.1 of \cite{2015A&A_Grazian_Schechter_functions} for redshift bins $z_n>5.5$.

\begin{figure}
    \centering
    \includegraphics[width=0.5\textwidth]{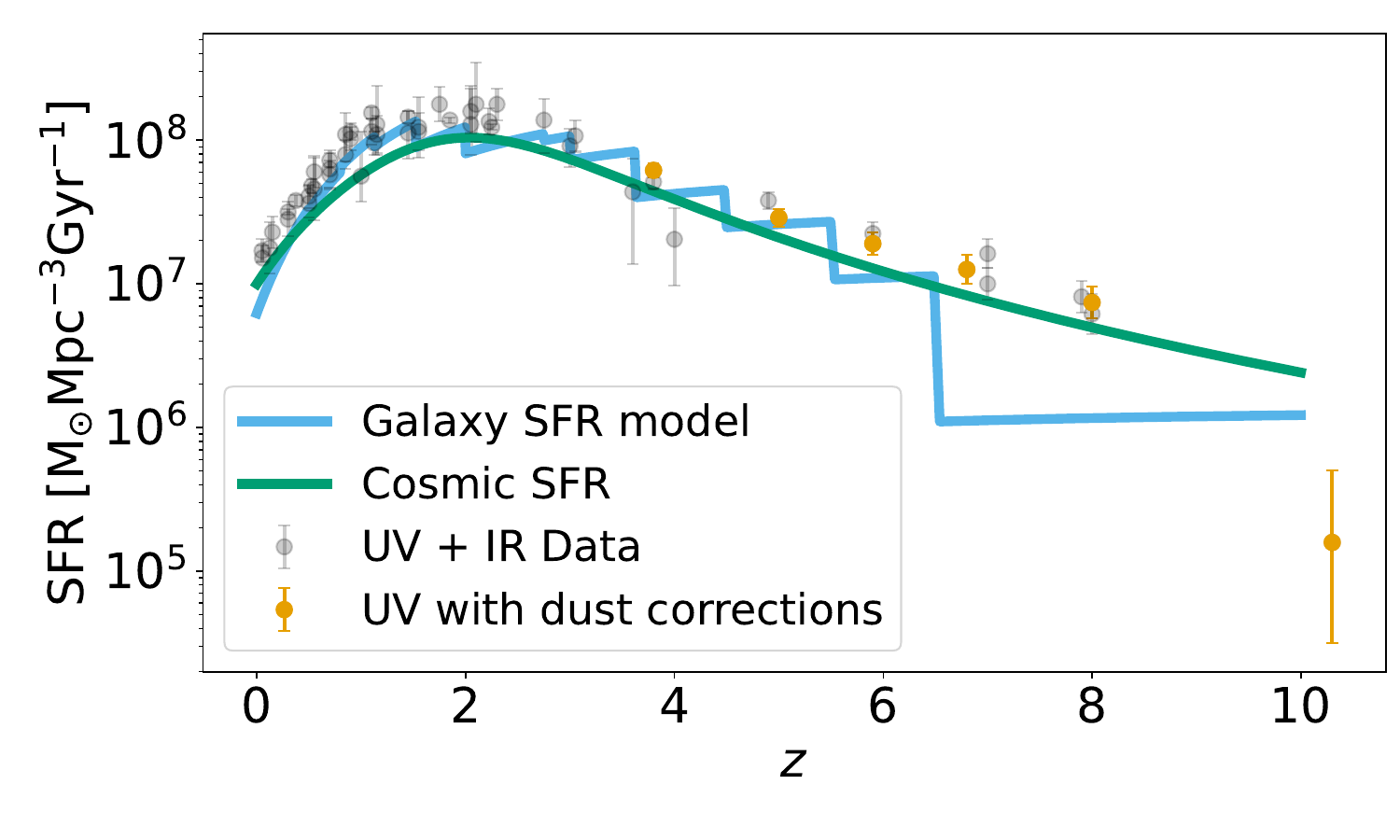}
    \caption{The cosmic SFR ( \citealt{Madau_Fargos_2014}; green line) and the galaxy SFR model \citep{2014ApJ_Speagle_GSFR} with galaxy density function of Eq.~\eqref{eq:Press_Schechter} (blue line), integrated over galaxy stellar mass according to the left-hand side of Eq.~\eqref{eq:cosmic_SFR_From_galaxy_SFR}. The grey circles with error bars represent ultra-violet and infra-red data from Table~1 of \protect\cite{MadauDickinson2014} and the orange circles with error bars represent dust-corrected ultra-violet data from Table 4 of \citet{2014ApJ_Bouwens_high_redshift_SFR}. This comparison illustrates the approximation in \equnref{eq:cosmic_SFR_From_galaxy_SFR}, which is deficient at redshifts $z>6.5$ (irrelevant to the present correlation method) due to limited observations and weakly constrained Press-Schechter function parameters.}
    \label{fig:galaxysfr}
\end{figure}

To model the SFR of individual galaxies of stellar mass $\mathcal{\tilde{M}}$, we use the star-forming galaxy main sequence SFR model $\bar{\psi}_{SF, \mathcal{\tilde{M}}}(t)$ (in $M_{\odot}\, \mathrm{yr}^{-1}$) as proposed by \citet{2014ApJ_Speagle_GSFR} (the ``preferred fit'' in their Table 9). The SFR accounts for accretion and outflow of stellar matter and assumes a galaxy stellar seed mass of $10^7 M_\odot$.
As illustrated in Fig.~\ref{fig:galaxysfr}, we verify that the galaxy SFR model, together with the galaxy density function, is consistent with the cosmic SFR $\bar{\rho}_{SF} (t)$ (e.g., as reported by \citet{Madau_Fargos_2014}), that is,
\begin{equation}
    \int^{\tilde{\mathcal{M}}_{\rm max}}_{\tilde{\mathcal{M}}_{\rm min}} \bar{\psi}_{SF, \mathcal{\tilde{M}}} (t)  N(\mathcal{\tilde{M}}, t) \mathrm{d}\mathcal{\tilde{M}} \simeq \bar{\rho}_{SF} (t), \label{eq:cosmic_SFR_From_galaxy_SFR}
\end{equation}
where $\mathcal{\tilde{M}}_{\rm min} = 8.5$ and $\mathcal{\tilde{M}}_{\rm max} = 12$ define the minimum and maximum galaxy stellar masses considered here. The integration limits are chosen such that the cosmic SFR at $z<6$ is captured accurately by the galaxy SFR model without large deviations at $z>6$. In this model, galaxies with stellar mass $10^{8}-10^{12} M_{\odot}$ dominate star formation in the Universe at $z<6$ (see also Fig.~\ref{fig:galaxysfr}). We verify that our results are converged for $z<6$ using a step size for numerical integration of $\leq 10^{0.5} M_{\odot}$. 

In each galaxy of stellar mass $\tilde{M}$, the BNS and rCCSNe rates (number of events per time) are calculated by the convolution of the galaxy SFR with the respective DTD as defined in \equnref{eq:delay_time_definition}, $D(t;t_{\mathrm{min}},b)_{\mathrm{BNS}} = \Theta(t-t_{\mathrm{min}})t^{b}$ and $D(t)_{\mathrm{rCCSNe}} = \delta(t)$:
 \begin{equation}
     R_{\mathrm{s}, \mathcal{\tilde{M}}}(t) = N_{\rm norm, s}  \int_{t_0}^{t} D(t-t';t_{\mathrm{min}},b)_{\mathrm{s}}  \bar{\psi}_{\mathrm{SF}, \mathcal{\tilde{M}}}(t') \mathrm{d}t'. \label{eq:galaxy_SFR_prop_eq}
 \end{equation}
 Here, $\mathrm{s}\in \{\mathrm{BNS,rCCSNe}\}$, $t_0 = t(z=10)$, and the DTD parameters $t_{\rm min},b$ of BNS mergers are free parameters. We assume the same effective production efficiency of BNS and rCCSNe events per stellar mass formed across galaxies of all masses and at all times. The galaxy-specific event rates are then normalized by the constraint that the galaxy-integrated event rates must match the local observed volumetric rates $R_{\mathrm{s,z0}}$ at $z=0$ (number of events per time per local comoving volume),
\begin{equation}
    R_{\mathrm{s,z0}} = \int_{t_0}^{t(z=0)} \mathrm{d}t' \int_{\tilde{\mathcal{M}}_{\rm min}}^{\tilde{\mathcal{M}}_{\rm max}} \mathrm{d} \mathcal{\tilde{M}} \, N(\mathcal{\tilde{M}},t') R_{\mathrm{s}, \mathcal{\tilde{M}}}(t).
\end{equation}
 
For the observed current rates of rCCSNe, we take $R_{\mathrm{rCCSNe,z0}}\approx 4-20$\,Gpc$^{-3}$yr$^{-1}$ at $z=0$, motivated by collapsars as the progenitors of long GRBs \citep{2019Nature_Siegel}. The current observed BNS merger rate at $z=0$ is $R_{\mathrm{BNS,z0}}\approx 10-250$ \, Gpc$^{-3}$yr$^{-1}$ \citep{2025_GWTC4_r_and_p}.

The rate of Eu production $C_\mathrm{s,\mathcal{\tilde{M}}}(t)$ (i.e., the mass synthesized per time) by BNS and rCCSNe in the ISM of a given galaxy is equal to their respective galaxy-specific rates multiplied by their corresponding average yield per event, $C_\mathrm{s,\mathcal{\tilde{M}}}(t) = m_{\mathrm{Eu,s}} R_{\mathrm{s}, \mathcal{\tilde{M}}}(t)$. In addition to the DTD parameters, the production of Eu is controlled by the effective parameter
 \begin{equation}
 M_{\mathrm{s}} = m_{\mathrm{Eu,s}} N_{\rm norm, s}  \label{eq:M_s},  
 \end{equation}
which represents the mass of Eu produced per stellar mass formed for a given source (BNS, rCCSNe). Mass-loss through galactic outflows and star formation scales with the SFR and is taken into account via a simple loss term such that the mass of Eu due to a source s in a given galaxy with logarithmic stellar mass $\tilde{\mathcal{M}}$ evolves according to
\begin{equation}
    \frac{\mathrm{d} M_{\mathrm{Eu},\mathcal{\tilde{M}}, \mathrm{s}}(t)}{\mathrm{d}t} =  C_\mathrm{s,\mathcal{\tilde{M}}}(t)-M_{\mathrm{Eu},\mathcal{\tilde{M}}, \mathrm{s}}(t) \ f_{\mathcal{\tilde{M}}}(t) \ \label{eq:galaxy_Eu_rate}.
\end{equation}
The mass-loss factor is given by $f_{\mathcal{\tilde{M}}}(t) =  (1+0.25)\times [\bar{\psi}_{\mathrm{SF},\mathcal{\tilde{M}}}(t)/M_{\mathrm{ISM},\mathcal{\tilde{M}}}(t)]$ (see \citealt{2019Nature_Siegel}), where $M_{\mathrm{ISM},\mathcal{\tilde{M}}}$ denotes the ISM mass of a galaxy of stellar mass $\mathcal{\tilde{M}}$, calculated as detailed below.

Next, we model the hydrogen content of individual galaxies. The mass $M_{\rm HI,\mathcal{\tilde{M}},z0}$ of neutral hydrogen HI of a galaxy of logarithmic stellar mass $\mathcal{\tilde{M}}$ and halo mass $\mathcal{\tilde{M}_H}$ at $z=0$ is taken from \citet{2019_Obuljen_HI_content_in_galaxy} as
\begin{equation}
M_{\rm HI,\mathcal{\tilde{M}},z0} = M_0 \left[\frac{\mathcal{\tilde{M}_H}(\tilde{\mathcal{M}})}{\mathcal{M}_{\rm min, HI}}\right]^{\alpha_{\rm HI}}\exp\left[-\frac{\mathcal{M}_{\rm min, HI}}{\mathcal{\tilde{M}_H}(\tilde{\mathcal{M}})}\right],
\end{equation}

where $\log_{10} \left(M_0/M_{\odot}\right) = 9.44^{+0.31}_{-0.39} \ h^{-1}$, $\log_{10} \left(\mathcal{M}_{\rm min, HI}/M_{\odot}\right) = 11.18^{+0.28}_{-0.35} \ h^{-1}$, and $\alpha_{\rm HI} = 0.48 \pm 0.08$ \citep{2019_Obuljen_HI_content_in_galaxy}. The factor $h = 0.7$ allows for corrections due to the uncertainty in the Hubble constant. The dark matter halo masses $\mathcal{\tilde{M}_H}$ are related to the galaxy stellar masses $\mathcal{\tilde{M}}$ via the redshift-dependent stellar halo mass (SHM) relation $\mathcal{S}(z)$ as given in \cite{2013_Moster_galactic_SFR}, i.e., $\mathcal{\tilde{M}_H}(z) = \mathcal{S}(z)\mathcal{\tilde{M}}$. Henceforth, we substitute the halo masses with the corresponding stellar mass. The Kennicutt-Schmidt relation then provides the evolution of the neutral hydrogen mass as a function of redshift as\footnote{the Kennicutt-Schmidt relation generally refers to the mass of the ISM. However, hydrogen is the dominant part ($\approx\!75\%$) of the ISM gas.} 
\begin{equation}
    M_{\mathrm{HI},\mathcal{\tilde{M}}} (t) =  M_{\rm HI,\mathcal{\tilde{M}},z0} \ \left[\frac{\bar{\psi}_{\mathrm{SF},\mathcal{\tilde{M}}}(t)}{ \bar{\psi}_{\mathrm{SF},\mathcal{\tilde{M}}}(z=0)}\right]^{\frac{1}{1.4}}.
\end{equation}
 Molecular hydrogen H$_2$ being less abundant and more difficult to observe than HI is relatively poorly constrained. We approximate its ISM mass $M_{\mathrm{H}_2,\mathcal{\tilde{M}}}$ by $30\%$ of the atomic hydrogen, following the average molecular-to-atomic hydrogen ratio found by \citet{2011_cold_gass_molecular_hydrogen}.
 
 Taking 75\% of the ISM mass to be hydrogen, we therefore calculate the ISM mass of a galaxy of stellar mass $\mathcal{\tilde{M}}$ as
\begin{equation}
    M_{\mathrm{ISM},\mathcal{\tilde{M}}}(t) = \frac{4}{3}\left[M_{\mathrm{HI},\mathcal{\tilde{M}}} (t) + M_{\mathrm{H_2},\mathcal{\tilde{M}}} (t) \right] 
    = 1.73 \times  M_{\mathrm{HI},\mathcal{\tilde{M}}} (t) . \label{eq:ISM_mass_per_galaxy}
\end{equation}
The galaxy-specific number of europium and hydrogen atoms in the ISM are given by 
\begin{equation}
     N_{\mathrm{Eu},\mathcal{\tilde{M}},\mathrm{s}}(t) = \frac{M_{\mathrm{Eu},\mathcal{\tilde{M}}, \mathrm{s}}}{m_{\mathrm{Eu}}}, \mskip40mu 
    N_{\mathrm{H},\mathcal{\tilde{M}}}(t) =  \frac{M_{\mathrm{HI},\mathcal{\tilde{M}}} (t)}{m_{\mathrm{HI}}} +   \frac{2\ M_{\mathrm{H}_2,\mathcal{\tilde{M}}(t)}}{m_{\mathrm{H}_2}}, 
     \label{eq:BNS_abund}
\end{equation}
where $m_{\mathrm Eu} = 151.96$\,amu, $m_{\mathrm{HI}} = 1.00$\,amu and $m_{\mathrm{H}_2} = 2.01$\,amu are the corresponding mean atomic weights. 
The average cosmic ratio of the number of europium to the number of hydrogen atoms at time $t$ from BNS and rCCSNe is calculated as the average of the galaxy-specific ratio weighted by the galaxy density function over all galaxy masses and total comoving volume $V_c$,

\begin{eqnarray}
     \frac{N_{\mathrm{Eu,s}}}{N_{\rm H}}(t)&=&\frac{1}{\int_{V(t = t_{0})}^{V(t)} \mathrm{d}V_c (t') \int_{\mathcal{\tilde{M}}_{\rm min}}^{\mathcal{\tilde{M}}_{\rm max}} \mathrm{d}\mathcal{\tilde{M}}\ \frac{\mathrm{d}^2n}{\mathrm{d}\mathcal{\tilde{M}}\mathrm{d}V_c}}\nonumber\\
     &\times& \int_{V(t = t_{0})}^{V(t)} \mathrm{d}V_c (t') \int_{\mathcal{\tilde{M}}_{\rm min}}^{\mathcal{\tilde{M}}_{\rm max}} \mathrm{d}\mathcal{\tilde{M}}\ \frac{\mathrm{d}^2n}{\mathrm{d}\mathcal{\tilde{M}}\mathrm{d}V_c} \frac{N_{\mathrm{Eu,\mathcal{\tilde{M}},s}}(t)}{N_{\rm H}(t)},\nonumber\\
     \\
     &=& \frac{1}{\int_{t_{0}}^{t} \mathrm{d}t' \frac{\mathrm{d}z}{\mathrm{d}t'}\frac{\mathrm{d}V_c (t')}{\mathrm{d}z} \int_{\mathcal{\tilde{M}}_{\rm min}}^{\mathcal{\tilde{M}}_{\rm max}} \mathrm{d}\mathcal{\tilde{M}}\ \frac{\mathrm{d}^2n}{\mathrm{d}\mathcal{\tilde{M}}\mathrm{d}V_c}}\nonumber\\
     &\times& \int_{t_{0}}^{t} \mathrm{d}t' \frac{\mathrm{d}z}{\mathrm{d}t'}  \frac{\mathrm{d}V_c}{\mathrm{d}z} \int_{\mathcal{\tilde{M}}_{\rm min}}^{\mathcal{\tilde{M}}_{\rm max}} \mathrm{d}\mathcal{\tilde{M}}\, N(\mathcal{\tilde{M}}, t')  \frac{N_{\mathrm{Eu,\mathcal{\tilde{M}},s}}(t)}{N_{\rm H}(t)}.\nonumber\\ 
\end{eqnarray} 

We then define our r-process abundance proxy as the corresponding abundance bracket $[\mathrm{Eu/H}]$, shifted by a constant offset $+5$ to avoid pathological behaviour of the correlation function when $[\mathrm{Eu/H}]$ approaches the solar value $[\mathrm{Eu/H}]=0$:
\begin{align}
    {\mathrm{Ab}^{\mathrm{Eu}}}_{\mathrm{model}}(t) =& \log_{10}\left[ \left(\frac{N_{\mathrm{Eu,BNS}}}{N_{\rm H}}(t) + \frac{N_{\mathrm{Eu,rCCSNe}}}{N_{\rm H}}(t)\right)\right. \nonumber\\
    &\qquad\qquad\qquad\qquad\qquad\qquad\quad\quad\times \left.\frac{N_{\mathrm{H,\odot}}}{N_{\mathrm{Eu,\odot}}}\right] + 5.  \label{eq:final_model}
\end{align}
We define the `fractional contribution' of BNS as
\begin{equation}
 F_{\rm BNS}(z(t)) = \frac{\mathrm{Ab^{Eu}}_{\mathrm{BNS}}(t)}{\mathrm{Ab^{Eu}}_{\mathrm{BNS}}(t) + \mathrm{Ab^{Eu}}_{\mathrm{rCCSNe}}(t)} , \label{eq:F_BNS_reln2}
\end{equation}
where $\mathrm{Ab^{Eu}}_{\mathrm{s}}(t) = \log_{10}\left[\frac{N_{\mathrm{Eu,s}}}{N_{\rm H}}(t) \times \frac{N_{\mathrm{H,\odot}}}{N_{\mathrm{Eu,\odot}}}\right] + 5$.

 The fractional contribution of BNS mergers to the r-process as a function of redshift (specifically at $z=0$, i.e., $F_{\rm{BNS,z0}} \equiv F_{\rm BNS}(z = 0)$) can be constrained jointly with the DTD parameters $t_{\rm min}, b$ with future observations of GWs and r-process abundances using this chemical evolution model. The rate-yield values $M_\mathrm{s}$ (\equnref{eq:M_s}) currently have large uncertainties; here, we treat these as free input parameters within bounds imposed by observational uncertainties in the rates (see above) and expectations of the range of mean ejecta yields of these events. We bracket the per-event Eu yield of BNS mergers by $m_{\mathrm{Eu,BNS}} = 10^{-5}-3\times10^{-4} M_{\odot}$, motivated by kilonova observations and numerical simulations of BNS mergers assuming a solar abundance pattern (e.g., \citealt{rastinejad_uniform_2025,kruger_estimates_2020}). For rCCSNe, we bracket the per-event yield by $m_{\mathrm{Eu,rCCSNe}} = 10^{-5} - 10^{-3}M_{\odot}$, based on numerical simulations of magnetorotational supernovae and collapsar accretion disks \citep{halevi_r-process_2018,2019Nature_Siegel}. For our Fisher analysis in \secref{sec:fisher Analysis}, different possible astrophysical scenarios are considered by making a fixed choice of model parameters $\theta = (t_{\mathrm{min}},b,F_{\rm{BNS,z0}})$. For fixed $\theta$, we adjust the two parameters $M_\mathrm{s}$ via root-finding such that simultaneously the desired value of $F_{\rm{BNS,z0}}$ is achieved and that the observed $z\approx 0$ Eu abundance in Milky Way-type galaxies ($\tilde{\mathcal{M}}=10$) of $[\mathrm{Eu/H}]=0.2\pm0.06$ \citep{2018_JINA_database, SAGA_database} is recovered.

\section{The Fisher Matrix}
\label{app:Fisher_derivation}

For a generic Gaussian log-likelihood,
\begin{equation}
    \mathcal{L^{\rm gen}} = \ln \det \mathcal{C} + (\mathcal{D}-\mathcal{\mu}(\beta))\mathcal{C}^{-1}(\mathcal{D}-\mathcal{\mu}(\beta))^T + \ln (2\pi), 
\end{equation}
where $\mathcal{C}$ is the data covariance matrix, $\mathcal{D}$ is the data matrix, and $\mathcal{\mu}$ is the expected data mean. The Fisher matrix is defined as the expectation of the curvature of the log-likelihood across multiple datasets with different noise realizations. The Fisher matrix elements $\mathds{F}^{\rm gen}_{pq}$ are

\begin{align}
    \mathds{F}^{\rm gen}_{pq}&= \bigg\langle\pdv{\mathcal{L}^{\rm gen}}{\beta_p,\beta_q}\bigg\rangle_{\beta_{\rm ML}}\nonumber\\ 
    &= \frac{1}{2}\rm{Tr}\left[\mathcal{C}^{-1}\pdv{\mathcal{C}}{\beta_p}\mathcal{C}^{-1}\pdv{\mathcal{C}}{\beta_q}
    + \mathcal{C}^{-1}\pdv{\mathcal{\mu(\beta)}}{\beta_p}\pdv{\mathcal{\mu^T(\beta)}}{\beta_q}\right. \nonumber\\
    &\qquad\qquad\qquad\qquad\qquad\qquad+ \left.\mathcal{C}^{-1}\pdv{\mathcal{\mu(\beta)}}{\beta_q}\pdv{\mathcal{\mu^T(\beta)}}{\beta_p}\right], 
    \label{eq:Fisher_matrix_general_form}
\end{align}

where $\beta_{\rm ML}$ is the maximum likelihood parameter set and $\beta_p,\beta_q \in \beta_{\rm ML}$ (see \citealt{1996ApJ_Vogeley_Fisher_Derivation}, their Appendix A, and \citealt{Tegmark:1996bz}, their Eq.~(15)).

In this work, the covariance matrices $\mathcal{C}=\mathds{C}_k$ and covariance functions $C_k$ (refer to \secref{sec:Framework}) are weakly dependent on the model parameters $\beta=\theta$ as compared to the expected data means $\mu = \mathcal{L}_k$ (refer to \secref{sec:fisher Analysis} for the definition of $\mathcal{L}_k$). For example, for counting statistical noise such as $\sigma_{\hat{N}_{\rm GW}} = \sqrt{\hat{N}_{\rm GW}} \propto \sqrt{N_{\rm GW,model}(\theta)}$ (see \secref{sec:Framework}), the dependence of counting uncertainty on $\theta$, i.e., $\pdv{\sigma_{\hat{N}_{\rm GW}}}{\theta_p}\propto\frac{1}{2\sqrt{N_{\rm GW,model}(\theta)}}\pdv{N_{\rm GW,model}}{\theta_p}$, where $\theta_p \in \theta$, is much smaller than the expected mean's dependence on $\theta$, i.e., $\partial N_{\rm GW,model}/\partial \theta_p$, when the number of GW observations is large. Similarly, redshift and abundance measurement uncertainties are assumed to be dependent mainly on the detectors and hence weakly dependent on the model parameters. Therefore, the covariance matrices that quantify measurement uncertainty are assumed to be independent of model parameters and are derived from mock data. 
Consequently, the first term in \equnref{eq:Fisher_matrix_general_form} vanishes. Furthermore, the remaining two terms are identical, owing to the diagonal form of the covariance matrix (see \secref{sec:Framework}), and hence add up to give the form of the Fisher matrix of \equnref{eq:derived_Fisher_matrix_form}.

\section{Correlation Trajectories for Abundance Proxies without the Offset}
\label{app:coorelation_without_offset}

\begin{figure}
    \centering
    \includegraphics[width=0.5\textwidth]{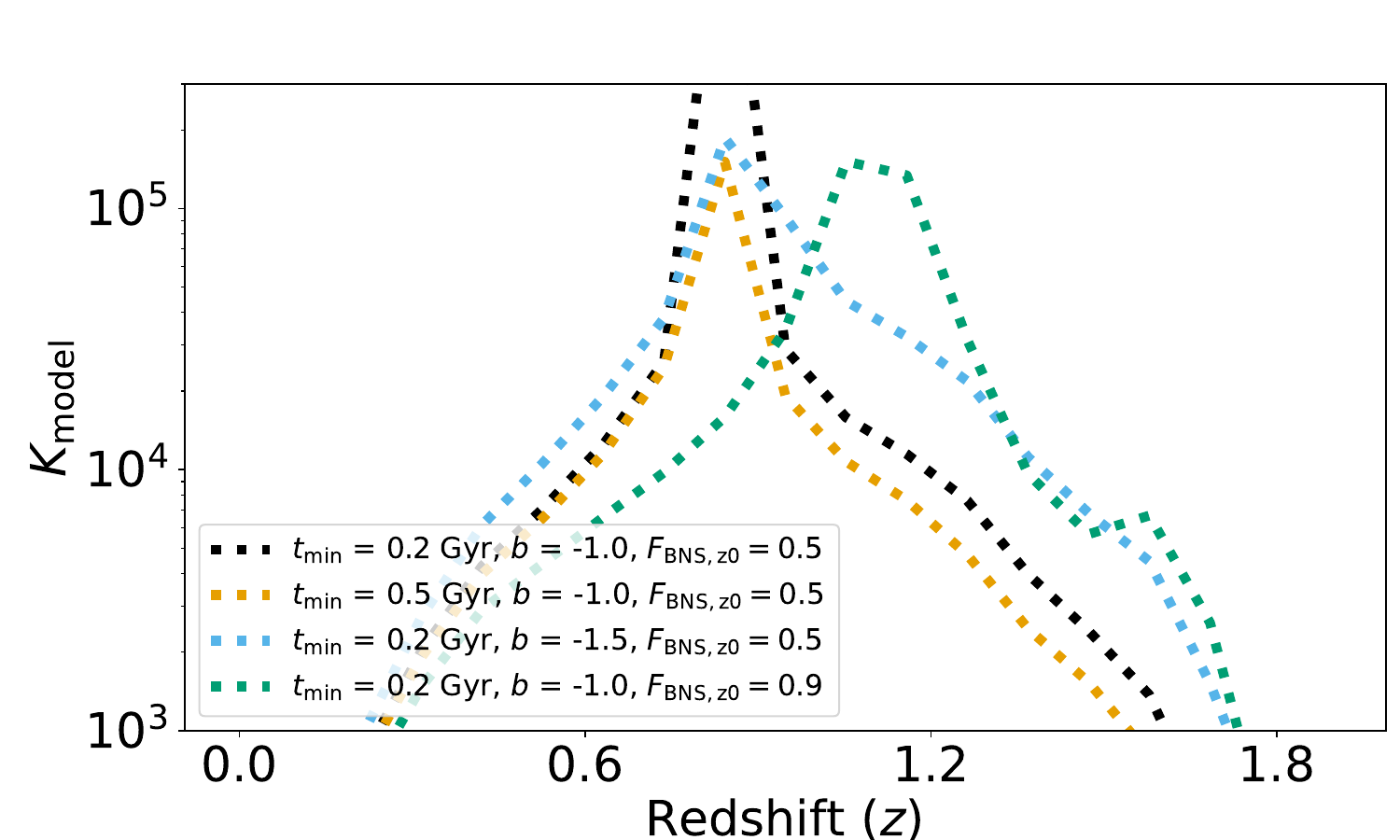}
    \caption{Analogous to the bottom panel of \figref{fig:corr_demo_plot}, the modeled correlation functions $K_{\rm model}$ as a function of redshift for different DTD parameters $t_{\rm min}$, $b$ and fractional contributions $F_{\rm{BNS,z0}}$ of BNS to cosmic r-process enrichment, except without the offset $+5$ to the abundance proxy $\mathrm{Ab^{Eu}}$ as described in \secref{sec:GCE_modelling}. Visually, the correlation function is sensitive to $F_{\rm{BNS,z0}}$, but $K_{\rm model}$ diverges as $[\mathrm{Eu/H}]$ approaches the solar value. This precludes application of a Fisher analysis in this redshift range.
    }
    \label{fig:correlation_without_offset}
\end{figure}

The correlation trajectories presented in \figref{fig:corr_demo_plot} visually appear to have less sensitivity to $F_{\rm BNS,z0}$ compared to the DTD parameters $t_{\rm min}$ and $b$. However, the uncertainty estimates of the Fisher forecasts in Table~\ref{tab:fisher_forecasts} show higher significance or signal-to-noise ratio for $F_{\rm BNS,z0}$ compared to the DTD parameters. Below we argue that the apparent smaller sensitivity in \figref{fig:corr_demo_plot} is due to the constant offset $+5$ introduced in the abundance proxy $\mathrm{Ab^{Eu}}$ as described in \secref{sec:GCE_modelling}, and show that the Fisher estimates are unaffected by this constant offset. For comparison, \figref{fig:correlation_without_offset} shows correlation functions analogous to \figref{fig:corr_demo_plot}, but when removing the constant offset to the abundances. The correlation function do show strong variations depending on the assumed value of $F_{\rm BNS,z0}$. In particular, the redshift at which the correlation function `peaks' (formally diverges) strongly depends on $F_{\rm BNS,z0}$. However, Fisher analyses cannot be applied in regions of mathematical divergences, which motivates applying a constant offset to the abundances to shift a divergence out of the range of interest.

The Fisher information matrix element corresponding to a parameter $\xi$ for an observable $\mathds{O}$ with an absolute uncertainty $\sigma_{\mathds{O}}$ is given by
\begin{equation}
    \mathds{F}_{\xi\xi} \sim \frac{1}{\sigma^2_{\mathds{O}}} \left(\pdv{\mathds{O}}{\xi}\right)^2  
    = \left(\frac{1}{\xi}\right)^2 \left(\frac{\mathds{O}}{\sigma_{\mathds{O}}}\right)^2 \left(\pdv{\mathds{\ln O}}{\ln \xi}\right)^2.
\end{equation}
Thus the relative uncertainty estimate from the Cramer-Rao bound on $\xi$ 
scales as
\begin{equation}
    \frac{\sigma_{\xi}}{\xi} \sim \left(\frac{\mathds{O}}{\sigma_{\mathds{O}}}\right)^{-1} \left(\pdv{\mathds{\ln O}}{\ln \xi}\right)^{-1} .\label{eq:Fisher_fractional_uncertainty}
\end{equation}
For the likelihood term $L_1$ in \equnref{eq:general_likelihood_gaussian_proposed}, $\mathds{O}\equiv \hat{K}$, $\xi \equiv F_{\rm BNS,z0}$, and since $\hat{K} \propto [\mathrm{\hat{Ab^{Eu}}}(z,F_{\rm BNS,z0})]^{-1}$, the fractional sensitivity 
\begin{equation}
    \left(\pdv{\mathds{\ln O}}{\ln \xi}\right)^{-1} \equiv \left(\pdv{\mathds{\ln \hat{K}}}{\ln F_{\rm BNS,z0}}\right)^{-1} = \left(\pdv{\mathds{\ln \hat{K}}}{\ln \mathrm{\hat{Ab^{Eu}}}}\pdv{\ln \mathrm{\hat{Ab^{Eu}}}}{\ln F_{\rm BNS,z0}}\right)^{-1}
\end{equation}
decreases by a factor $\mathrm{\hat{Ab}^{Eu}}/(\mathrm{\hat{Ab}^{Eu}}+c)$, for a constant linear offset $c$ (corresponding to a factor of $\gtrsim\! 6$ for $\hat{\mathrm{Ab}}^{\rm Eu}\sim 0.1-1$ and $c=5$). 
However, the linear offset also results in an increase by a factor of  $(\mathrm{\hat{Ab}^{Eu}}+c) / \mathrm{\hat{Ab}^{Eu}}$ in the significance of the correlation measurement ($\hat{K}/\sigma_{\hat{K}-\mathrm{\hat{Ab}^{Eu}}}\equiv \mathds{O}/\sigma_{\mathds{O}}$; see \equnref{eq:abundance_error}). 
The relative uncertainty of the estimate $\xi = F_{\rm BNS,z0}$ remains unaffected (\equnref{eq:Fisher_fractional_uncertainty}). Similarly, for the likelihood term $L_2$ in \equnref{eq:general_likelihood_gaussian_proposed}, the fractional sensitivity of the abundances $\mathrm{\hat{Ab}^{Eu}}$ to $F_{\rm BNS,z0}$ decreases and the significance of $\mathrm{\hat{Ab}^{Eu}}$ increases by the inverse factor due to the constant offset $c$, keeping the relative Fisher uncertainty estimate invariant.

\section{Mock Gravitational-Wave Data Generation For Calculation Of Luminosity Distance Uncertainties}
\label{sec:Mock_GW_catalogue}

 To estimate the luminosity distance uncertainty in the dark-siren case (absence of EM counterparts providing a direct redshift measurement), we create a mock GW catalogue of projected future observations by the 3G detectors CE (\citealt{2019_Cosmic_explorer_white_paper}) and ET (\citealt{2020_Einstein_Telescope_white_paper}). 
 The detector sensitivity is defined by the noise curves of ET and CE available in the LIGO database\footnote{https://dcc.ligo.org/LIGO-T1500293/public}. The number of GW sources $N_{\rm GW}(z_n)$ per year of observation in a redshift bin $z_n$ for the detectors is calculated using \equnref{eq:N_GW}. The total number of GW sources across cosmic history is given by $T_{\rm GW} = \int_{z=0}^{10}\mathrm{d}z N_{\rm GW}(z) \approx \sum_{z_n} N_{\rm GW}(z_n)$. 
 
 The neutron star (NS) mass distribution in BNS is assumed to be a Gaussian with mean $\mu$ = 1.4 $M_{\odot}$ and standard deviation $\sigma = 0.2 M_{\odot}$, truncated at $1 \ M_{\odot}$ and $2.4 \ M_{\odot}$ \citep{2016_Ozel_DNS_population}. Taking one year of observation time, two sets $\mathds{M}_1$ and $\mathds{M}_2$ of NS masses with $T_{\rm GW}$ elements each are sampled from the NS mass distribution. Upon pairing the set elements $\mathcal{S}_1 \in \mathds{M}_1$, $\mathcal{S}_2 \in \mathds{M}_2$, we obtain $T_{\rm GW}$ BNS systems with chirp-mass distribution $\mathds{M}_{\rm c} = \{(\mathcal{S}_1 \mathcal{S}_2)^{3/5}/(\mathcal{S}_1 + \mathcal{S}_2)^{1/5}\}$ and mass-ratio distribution $\mathds{Q} = \{\mathcal{S}_2/\mathcal{S}_1\}$. The redshift distribution of these sources is given by $\mathcal{D} = N_{\rm GW}(z)/T_{\rm GW}$. We sample $T_{\rm GW}$ redshifts from this distribution, which are converted to luminosity distances assuming the standard \citet{2018_Planck_cosmological_params} cosmology.
 
 Using the chirp mass and luminosity distance samples, the signal-to-noise (SNR) ratio $\rho$ of these events for a combination of two CEs and one ET can be calculated using Eq.~(B1) of \citet{cosmology_dark_Simone_2021},
 \begin{equation}
     \rho = \rho^{\ast} \, \Theta \left( \frac{M_{\rm c}}{M_{\rm c}^{\ast}} \right)^{5/6} \left( \frac{d_{\rm L}^{\ast}}{d_{\rm L}} \right).\label{eq:snr_fit}
 \end{equation}
 Here, $M_{\rm c}$ is the chirp mass and $d_{\rm L}$ is the luminosity distance of the event, $M_{\rm c}^{\ast} = 1 M_{\odot}$, $d_{\rm L}^{\ast} = 1000 $\,Mpc, and $\Theta \in \mathds{U}(0,1)$, where $\mathds{U}$ is the uniform distribution. We derive $\rho_{\ast} \approx 81.08$ for CE and $\rho_{\ast} \approx 61.5$ for the ET using frequency domain waveforms generated by \texttt{PyCBC} \citep{Pycbc_software} for $1000$ of the $T_{\rm GW}$ BNS samples. We calculate the signal-to-noise ratio of these waveforms using expected detector sensitivities\footnote{\url{https://dcc.ligo.org/LIGO-T1500293/public}} and fit \equnref{eq:snr_fit} to calculate the values of $\rho_{\ast}$. Since waveform generation and subsequent SNR calculation are time-consuming processes, we calculate the SNR for all other samples in $\mathds{M}_{\rm c}$ separately for CE ($\rho_{\mathrm{CE}}$) and ET ($\rho_{\mathrm{ET}}$) using \equnref{eq:snr_fit}. We then add these in quadrature to obtain the total SNR $\rho_{\mathrm{total}} = \sqrt{\rm 2\rho_{\mathrm{CE}}^2 + \rho_{\mathrm{ET}}^2}$. Events with $\rho_{\mathrm{total}}> 12$ are considered detected sources.  The detected number of sources $\hat{N}_{\rm GW,detected}$ is used to calculate the selection function $\phi(z)$ of a detector configuration as $\phi(z) = \hat{N}_{\rm GW,detected}(z)/N_{\rm GW}(z)$. The uncertainty in luminosity distance measurements $\sigma_{d_{\rm L}}$ for the detected GW signals is computed using the quick posterior sampling method outlined in \citet{cosmology_dark_Simone_2021}, \citet{2019_future_percent_level_hubble_Expansion}, and \citet{2018_BH_merger_rate_evolution}. In this method, SNR-dependent noise is added to the chirp mass, mass ratio, and $\Theta$ samples in each redshift bin, and posteriors of luminosity distance are drawn by inverting \equnref{eq:snr_fit}. The covariance of the sampled luminosity distance posterior provides an estimate of luminosity distance uncertainty in different redshift bins.

\section{Sensitivity to fiducial assumptions}
\label{app:sensitivity}

\begin{table*}
    \centering
    \caption{Analogous to Tab.~\ref{tab:fisher_forecasts}, the $1\sigma$-uncertainty on estimated parameters for different detection scenarios: two CE detectors and one ET (2CE + ET), a single ET, and a single CE detector. We assume one year of observation time without EM counterparts and different astrophysical scenarios (varying assumptions of the true (underlying) DTD parameters $t_{\rm min}$, $b$ and the fractional BNS contribution to the current cosmic r-process abundances $F_{\rm{BNS,z0}}$). Numbers in parenthesis represent the significance or signal-to-noise ratio (SNR) of the result. }
    \begin{tabular}{lcccccr}
         \hline
         Detector Network  & $F_{\rm{BNS,z0}}$ & $t_{\mathrm{min}}$[Gyr] & $b$  & $\sqrt{\left(F^{-1}\right)_{F_{\rm{BNS,z0}}F_{\rm{BNS,z0}}}}$ (SNR)& $\sqrt{\left(F^{-1}\right)_{t_{\mathrm{min}}t_{\mathrm{min}}}}$ [Gyr] (SNR)& $\sqrt{\left(F^{-1}\right)_{bb}}$ (SNR) \\ 
         \hline
        2CE + ET & 0.1 & 0.2 & -1 & 0.03 ($0.3\times 10^1$)              &  0.3 (0.7) & 0.2 ($0.5\times 10^1$) \\
        
         2CE + ET & 0.5 & 0.2  & -1 & 0.03 ($2\times 10^1$) &  0.3 (0.7) & 0.2 ($0.5\times 10^1$) \\ 

          2CE + ET & 0.9  & 0.2 & -1 & 0.03 ($3\times 10^1$) & 0.3 (0.7) & 0.2 ($0.5\times 10^1$)\\ 
        
          2CE + ET & 0.5  & 0.2 & -2 & 0.05 ($1\times 10^1$) &  0.3 (0.7) & 0.5 ($0.4\times 10^1$) \\ 

         2CE + ET & 0.5  & 0.1 & -1 & 0.05 ($1\times 10^1$)  &  0.1 (1) & 0.2 ($0.5\times 10^1$) \\

         2CE + ET & 0.5 & 0.05 & -1 & 0.04 ($1\times10^1$)  &  0.08 (0.6) & 0.1 ($1\times 10^1$) \\
         
         2CE + ET & 0.5 & 1 & -1 & 0.03 ($2\times10^1$)  &  0.4 (3) & 0.1 ($1\times 10^1$) \\
         
         ET & 0.5  & 0.2 & -1 & 0.05 ($1\times10^1$)  &  1.3 (0.2) & 0.7 ($0.1\times 10^1$) \\
         
         CE & 0.5  & 0.2 & -1 & 0.04 ($1\times10^1$)  &  0.8 (0.3) & 0.4 ($0.3\times 10^1$) \\
         \hline
    \end{tabular}
    \label{tab:fisher_sensitivity}
\end{table*}

\begin{figure}
    \centering
    \includegraphics[width=0.5\textwidth]{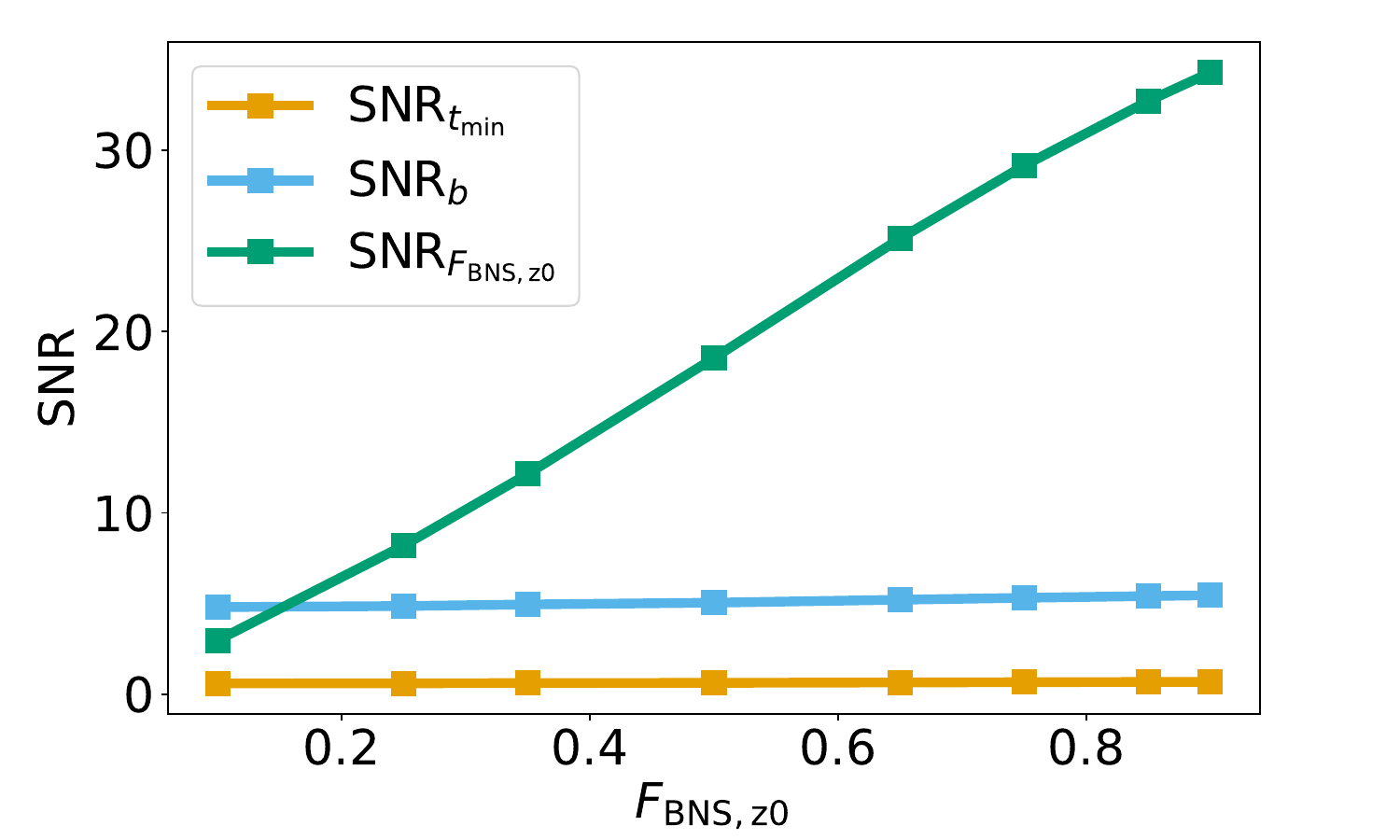}
    \caption{The signal-to-noise ratio (SNR) in estimating the DTD parameters $t_{\rm min}$ and $b$ as well as the fractional contribution of BNS to r-process enrichment $F_{\rm BNS,z0}$ for different astrophysical scenarios of the true underlying $F_{\rm{BNS,z0}}$, keeping all other parameters fixed to their fiducial assumptions, assuming one year of observation time without EM counterparts. The increasing SNR of $F_{\rm{BNS,z0}}$ with increasing $F_{\rm{BNS,z0}}$ is expected as the correlation between the GW and abundance data strengthens.}
    \label{fig:SNR_increasing_fbns}
\end{figure}

The Fisher forecasts presented in this work estimate the precision with which one may be able to retrieve the BNS DTD parameters and $F_{\rm{BNS,z0}}$ in the future. As the fiducial results are based on the assumed astrophysical scenario of $t_{\mathrm{min}} = 0.2$\,Gyr, $b = -1$, and $F_{\rm{BNS,z0}} = 0.5$ and a detector network of two CEs and one ET, here we explore the robustness of our main conclusions to varying astrophysical assumptions and the detector network. Table~\ref{tab:fisher_sensitivity} shows the estimated parameter uncertainties for different parameter choices and network configurations. 

The absolute uncertainty of measuring $F_{\mathrm{BNS,z0}}$ is largely insensitive to parameter variations (Rows $1-7$ of Table~\ref{tab:fisher_sensitivity}) and our fiducial case (Row $2$ of Table~\ref{tab:fisher_sensitivity}) remains representative. The significance or SNR of the fiducial case is also generally representative of parameter changes, but decreases with decreasing $F_{\mathrm{BNS,z0}}$ due to overall weaker correlations. Conversely, as $F_{\rm{BNS,z0}}$ increases, the correlation between the GW and abundance datasets increases, leading to increased precision in $F_{\rm{BNS,z0}}$ parameter estimation. This is illustrated with a more detailed analysis in \figref{fig:SNR_increasing_fbns}. 

The absolute uncertainty of the minimum delay time parameter $t_{\rm min}$ for the fiducial case constitutes a representative value across all parameter configurations with $t_{\rm min}\ll 1$\,Gyr and the 2CE + ET detector network (first six rows of Table.~\ref{tab:fisher_sensitivity}).

As $t_{\rm min}$ increases to $\gtrsim 1$\,Gyr (Row 7 of Table~\ref{tab:fisher_sensitivity}), the sensitivity of the average delay of BNS mergers (\equnref{eq:av_delay}) to variations in $t_{\rm min}$ increases strongly as discussed in \secref{subsec:Fisher_results}. Thus, our correlation method becomes more sensitive to this parameter, and the relative uncertainty decreases by a factor of a few relative to the fiducial case.

For the DTD power-law index $b$, the absolute uncertainty and SNR of the fiducial case are also representative of varying astrophysical scenarios for the 2CE+ET detection network, with exceptions probing two extremes. In one case, solely the absolute uncertainty increases by a factor of $\approx\!2$ when much steeper DTD with $b \lesssim -2$ are considered. For such steep DTDs, the average BNS delay time becomes similar to our fiducial choice of $t_{\rm min}$, increasing the parameter uncertainty. 
Conversely, for a very small minimum delay time of $t_{\rm min} = 0.05$\,Gyr, the absolute uncertainty of $b$ decreases and the significance (SNR) increases by a factor of $\approx\!2$, since the average BNS delay time is then more strongly controlled by $b$ (see \equnref{eq:av_delay}).

When considering a detector network comprised of a single CE or a single ET (Rows $8-9$ of Table~\ref{tab:fisher_sensitivity}), the uncertainty of estimating the DTD parameters $t_{\rm min}$ and $b$ increases by a factor of $3-4$ and the significance decreases accordingly. This is because a smaller detector network leads to a decreased number of detected GW sources $\hat{N}_{\rm GW}$ and a decrease in the radius of the detection horizon of BNS mergers. This decreases the sensitivity of the correlation $\hat{K}$  (which is proportional to $\hat{N}_{\rm GW}$) in the likelihood ($L_1$ in \equnref{eq:general_likelihood_gaussian_proposed}) to changes in DTD parameters. Since the likelihood term $L_1$ is the dominant term controlling the precision of the estimation of DTD parameters, as discussed in Sec.~\ref{subsec:data_covariance}, this decrease in sensitivity leads to an increase in uncertainty of the estimation of DTD parameters. However, the r-process abundance term $L_2$ in \equnref{eq:general_likelihood_gaussian_proposed} is not affected by such a decrease in the number of GW sources. Hence, the absolute uncertainty of estimating $F_{\rm BNS,z0}$, which is largely controlled by $L_2$, is similar to the fiducial case of the 2CE + ET configuration (Row 2 of Table~\ref{tab:fisher_sensitivity}).


\bsp	
\label{lastpage}
\end{document}